\pdfoutput=1
\documentclass[12pt,a4paper]{article}

\usepackage{ifthen} 
\usepackage{rotating}
\usepackage{dashrule}
\usepackage{array} 
\usepackage{dcolumn}
\usepackage{comment}
\usepackage{xcolor}

\usepackage{graphpap} 
\usepackage{makecell} 
\usepackage{rotating} 
\usepackage{tikz}
\usetikzlibrary{patterns.meta}
\usepackage{xcolor}
\usepackage{longtable} 

\usepackage{lscape}
\usepackage{multirow} 
\usepackage{mathrsfs}
\usepackage{mathtools} 
\usetikzlibrary{patterns}
\usepackage{nicefrac} 
\usepackage{transparent}
\usepackage{afterpage} 
\usepackage[normalem]{ulem}
\usepackage{cancel}

\usepackage{ifthen} 
\newboolean{pdflatex}
\setboolean{pdflatex}{true} 

\newboolean{articletitles}
\setboolean{articletitles}{true} 

\newboolean{uprightparticles}
\setboolean{uprightparticles}{true} 

\newboolean{inbibliography}
\setboolean{inbibliography}{false} 

\def\paperauthors{LHCb collaboration} 
\def\paperasciititle{Observation of the Bc -> J/psi pi+ pi0 decay} 
\def\papertitle{Observation of the~$\decay{\Bc}{\jpsi\pip\piz}$~decay} 
\def\paperkeywords{{High Energy Physics}, {LHCb}} 
\def\papercopyright{\the\year\ CERN for the benefit of the LHCb collaboration} 
\def\paperlicence{CC BY 4.0 licence}
\def\paperlicenceurl{https://creativecommons.org/licenses/by/4.0/}

\usepackage{wasysym}
\DeclareMathOperator*{\bigplus}{\scalerel*{+}{\sum}}
\usepackage{scalerel}

\usepackage[top=1in, bottom=1.25in, left=1in, right=1in]{geometry}

\usepackage{microtype}
\usepackage{lineno}  
\usepackage{xspace} 
\usepackage{caption} 

\usepackage{graphicx}  
\usepackage{color}
\usepackage{colortbl}
\graphicspath{{./figs/}} 

\usepackage{amsmath} 
\usepackage{amssymb}
\usepackage{amsfonts}
\usepackage{upgreek} 

\newcommand*\patchAmsMathEnvironmentForLineno[1]{%
\expandafter\let\csname old#1\expandafter\endcsname\csname #1\endcsname
\expandafter\let\csname oldend#1\expandafter\endcsname\csname
end#1\endcsname
 \renewenvironment{#1}%
   {\linenomath\csname old#1\endcsname}%
   {\csname oldend#1\endcsname\endlinenomath}%
}
\newcommand*\patchBothAmsMathEnvironmentsForLineno[1]{%
  \patchAmsMathEnvironmentForLineno{#1}%
  \patchAmsMathEnvironmentForLineno{#1*}%
}
\AtBeginDocument{%
\patchBothAmsMathEnvironmentsForLineno{equation}%
\patchBothAmsMathEnvironmentsForLineno{align}%
\patchBothAmsMathEnvironmentsForLineno{flalign}%
\patchBothAmsMathEnvironmentsForLineno{alignat}%
\patchBothAmsMathEnvironmentsForLineno{gather}%
\patchBothAmsMathEnvironmentsForLineno{multline}%
\patchBothAmsMathEnvironmentsForLineno{eqnarray}%
}

\usepackage{hyperxmp}

\usepackage[pdftex,
            pdfauthor={\paperauthors},
            pdftitle={\paperasciititle},
            pdfkeywords={\paperkeywords},
            pdfcopyright={Copyright (C) \papercopyright},
            pdflicenseurl={\paperlicenceurl}]{hyperref}

\usepackage[colorinlistoftodos,textsize=scriptsize]{todonotes}

\usepackage[bottom,flushmargin,hang,multiple]{footmisc}

\usepackage[all]{hypcap} 
\usepackage{comment}

\usepackage{xspace} 
\usepackage{upgreek}

\def\lhcb   {\mbox{LHCb}\xspace}

\def\MagUp {\mbox{\em Mag\kern -0.05em Up}\xspace}

\ifthenelse{\boolean{uprightparticles}}%
{
 
 \def\Pgamma      {\ensuremath{\upgamma}\xspace}

 \def\Pmu         {\ensuremath{\upmu}\xspace}                 
 \def\Pnu         {\ensuremath{\upnu}\xspace}                 
                  
 \def\Ppi         {\ensuremath{\uppi}\xspace}                 
                  
 \def\Prho        {\ensuremath{\uprho}\xspace}                 
                  
 \def\Ptau        {\ensuremath{\uptau}\xspace}

 \def\Ppsi        {\ensuremath{\uppsi}\xspace}

 \def\PDelta      {\ensuremath{\Delta}\xspace}                 
 \def\PXi         {\ensuremath{\Xi}\xspace}                 
 \def\PLambda     {\ensuremath{\Lambda}\xspace}                 
 \def\PSigma      {\ensuremath{\Sigma}\xspace}                 
 \def\POmega      {\ensuremath{\Omega}\xspace}                 
 \def\PUpsilon    {\ensuremath{\Upsilon}\xspace}
 \let\oldPi\Pi
 \def\PPi         {\ensuremath{\oldPi}\xspace}

 \def\PB      {\ensuremath{\mathrm{B}}\xspace}                 
                  
 \def\PD      {\ensuremath{\mathrm{D}}\xspace}

 \def\PJ      {\ensuremath{\mathrm{J}}\xspace}                 
 \def\PK      {\ensuremath{\mathrm{K}}\xspace}

 \def\PS      {\ensuremath{\mathrm{S}}\xspace}

 \def\PW      {\ensuremath{\mathrm{W}}\xspace}

 \def\Pa      {\ensuremath{\mathrm{a}}\xspace}                 
 \def\Pb      {\ensuremath{\mathrm{b}}\xspace}                 
 \def\Pc      {\ensuremath{\mathrm{c}}\xspace}                 
 \def\Pd      {\ensuremath{\mathrm{d}}\xspace}

 \def\Pi      {\ensuremath{\mathrm{i}}\xspace}

 \def\Pp      {\ensuremath{\mathrm{p}}\xspace}                 
 \def\Pq      {\ensuremath{\mathrm{q}}\xspace}                 
                  
 \def\Ps      {\ensuremath{\mathrm{s}}\xspace}                 
                  
 \def\Pu      {\ensuremath{\mathrm{u}}\xspace}

 \def\thebaroffset{0.0em}
}
{
 
 \def\Pgamma      {\ensuremath{\gamma}\xspace}

 \def\Pmu         {\ensuremath{\mu}\xspace}                 
 \def\Pnu         {\ensuremath{\nu}\xspace}                 
                  
 \def\Ppi         {\ensuremath{\pi}\xspace}                 
                  
 \def\Prho        {\ensuremath{\rho}\xspace}                 
                  
 \def\Ptau        {\ensuremath{\tau}\xspace}

 \def\Ppsi        {\ensuremath{\psi}\xspace}                 
                  
 \mathchardef\PDelta="7101
 \mathchardef\PXi="7104
 \mathchardef\PLambda="7103
 \mathchardef\PSigma="7106
 \mathchardef\POmega="710A
 \mathchardef\PUpsilon="7107
 \mathchardef\PPi="7105
                  
 \def\PB      {\ensuremath{B}\xspace}                 
                  
 \def\PD      {\ensuremath{D}\xspace}

 \def\PJ      {\ensuremath{J}\xspace}                 
 \def\PK      {\ensuremath{K}\xspace}

 \def\PS      {\ensuremath{S}\xspace}

 \def\PW      {\ensuremath{W}\xspace}

 \def\Pa      {\ensuremath{a}\xspace}                 
 \def\Pb      {\ensuremath{b}\xspace}                 
 \def\Pc      {\ensuremath{c}\xspace}                 
 \def\Pd      {\ensuremath{d}\xspace}

 \def\Pi      {\ensuremath{i}\xspace}

 \def\Pp      {\ensuremath{p}\xspace}                 
 \def\Pq      {\ensuremath{q}\xspace}                 
                  
 \def\Ps      {\ensuremath{s}\xspace}                 
                  
 \def\Pu      {\ensuremath{u}\xspace}

 \def\thebaroffset{0.18em}
}
\newcommand{\offsetoverline}[2][\thebaroffset]{\kern #1\overline{\kern -#1 #2}}%

\makeatletter
\ifcase \@ptsize \relax
  \newcommand{\miniscule}{\@setfontsize\miniscule{4}{5}}
\or
  \newcommand{\miniscule}{\@setfontsize\miniscule{5}{6}}
\or
  \newcommand{\miniscule}{\@setfontsize\miniscule{5}{6}}
\fi
\makeatother

\DeclareRobustCommand{\optbar}[1]{\shortstack{{\miniscule (\rule[.5ex]{1.25em}{.18mm})}
  \\ [-.7ex] $#1$}}

\def\mumu       {{\ensuremath{\Pmu^+\Pmu^-}}\xspace}

\def\neub       {{\ensuremath{\overline{\Pnu}}}\xspace}

\def\neutb      {{\ensuremath{\neub_\tau}}\xspace}

\def\g      {{\ensuremath{\Pgamma}}\xspace}

\def\Wp     {{\ensuremath{\PW^+}}\xspace}

\def\quark     {{\ensuremath{\Pq}}\xspace}
\def\quarkbar  {{\ensuremath{\overline \quark}}\xspace}

\def\uquark    {{\ensuremath{\Pu}}\xspace}

\def\dquark    {{\ensuremath{\Pd}}\xspace}
\def\dquarkbar {{\ensuremath{\overline \dquark}}\xspace}

\def\squark    {{\ensuremath{\Ps}}\xspace}

\def\cquark    {{\ensuremath{\Pc}}\xspace}
\def\cquarkbar {{\ensuremath{\overline \cquark}}\xspace}

\def\bquark    {{\ensuremath{\Pb}}\xspace}
\def\bquarkbar {{\ensuremath{\overline \bquark}}\xspace}

\def\pion   {{\ensuremath{\Ppi}}\xspace}
\def\piz    {{\ensuremath{\pion^0}}\xspace}
\def\pip    {{\ensuremath{\pion^+}}\xspace}
\def\pim    {{\ensuremath{\pion^-}}\xspace}

\def\rhomeson {{\ensuremath{\Prho}}\xspace}

\def\rhop     {{\ensuremath{\rhomeson^+}}\xspace}

\def\kaon    {{\ensuremath{\PK}}\xspace}

\def\KorKbar {\kern \thebaroffset\optbar{\kern -\thebaroffset \PK}{}\xspace}

\def\Kp      {{\ensuremath{\kaon^+}}\xspace}
\def\Km      {{\ensuremath{\kaon^-}}\xspace}

\def\KS      {{\ensuremath{\kaon^0_{\mathrm{S}}}}\xspace}

\def\Kstarp  {{\ensuremath{\kaon^{*+}}}\xspace}

\def\D       {{\ensuremath{\PD}}\xspace}

\def\DorDbar {\kern \thebaroffset\optbar{\kern -\thebaroffset \PD}\xspace}
\def\Dz      {{\ensuremath{\D^0}}\xspace}

\def\Dp      {{\ensuremath{\D^+}}\xspace}
\def\Dm      {{\ensuremath{\D^-}}\xspace}

\def\DpDm    {\ensuremath{\Dp {\kern -0.16em \Dm}}\xspace}

\def\Dstarp  {{\ensuremath{\D^{*+}}}\xspace}

\def\B       {{\ensuremath{\PB}}\xspace}

\def\BorBbar {\kern \thebaroffset\optbar{\kern -\thebaroffset \PB}\xspace}

\def\Bd      {{\ensuremath{\B^0}}\xspace}

\def\BdorBdbar {\kern \thebaroffset\optbar{\kern -\thebaroffset \Bd}\xspace}
\def\Bu      {{\ensuremath{\B^+}}\xspace}

\def\Bp      {{\ensuremath{\Bu}}\xspace}

\def\Bs      {{\ensuremath{\B^0_\squark}}\xspace}

\def\BsorBsbar {\kern \thebaroffset\optbar{\kern -\thebaroffset \Bs}\xspace}
\def\Bc      {{\ensuremath{\B_\cquark^+}}\xspace}
\def\Bcp     {{\ensuremath{\B_\cquark^+}}\xspace}
\def\Bcm     {{\ensuremath{\B_\cquark^-}}\xspace}

\def\jpsi     {{\ensuremath{{\PJ\mskip -3mu/\mskip -2mu\Ppsi}}}\xspace}
\def\psitwos  {{\ensuremath{\Ppsi{(2\PS)}}}\xspace}

\def\Y#1S{\ensuremath{\PUpsilon{(#1S)}}\xspace}

\def\proton      {{\ensuremath{\Pp}}\xspace}

\def\LorLbar     {\kern \thebaroffset\optbar{\kern -\thebaroffset \PLambda}\xspace}

\def\BF         {{\ensuremath{\mathcal{B}}}\xspace}
\def\BR         {\BF}

\newcommand{\decay}[2]{\ensuremath{#1\!\to #2}\xspace} 

\def\to                 {\ensuremath{\rightarrow}\xspace}

\def\AT#1     {\ensuremath{A_{\mathrm{T}}^{#1}}\xspace}           

\def\C#1      {\ensuremath{\mathcal{C}_{#1}}\xspace}                       
\def\Cp#1     {\ensuremath{\mathcal{C}_{#1}^{'}}\xspace}                    
\def\Ceff#1   {\ensuremath{\mathcal{C}_{#1}^{\mathrm{(eff)}}}\xspace}        
\def\Cpeff#1  {\ensuremath{\mathcal{C}_{#1}^{'\mathrm{(eff)}}}\xspace}       
\def\Ope#1    {\ensuremath{\mathcal{O}_{#1}}\xspace}                       
\def\Opep#1   {\ensuremath{\mathcal{O}_{#1}^{'}}\xspace}                    

\newcommand{\nospaceunit}[1]{\ensuremath{\text{#1}}}       
\newcommand{\aunit}[1]{\ensuremath{\text{\,#1}}}       

\newcommand{\tev}{\aunit{Te\kern -0.1em V}\xspace}
\newcommand{\gev}{\aunit{Ge\kern -0.1em V}\xspace}
\newcommand{\mev}{\aunit{Me\kern -0.1em V}\xspace}
\newcommand{\kev}{\aunit{ke\kern -0.1em V}\xspace}
\newcommand{\ev}{\aunit{e\kern -0.1em V}\xspace}
 
\newcommand{\mevc}{\ensuremath{\aunit{Me\kern -0.1em V\!/}c}\xspace}
\newcommand{\gevc}{\ensuremath{\aunit{Ge\kern -0.1em V\!/}c}\xspace}
\newcommand{\mevcc}{\ensuremath{\aunit{Me\kern -0.1em V\!/}c^2}\xspace}
\newcommand{\gevcc}{\ensuremath{\aunit{Ge\kern -0.1em V\!/}c^2}\xspace}

\def\mm   {\aunit{mm}\xspace}

\def\mum  {\ensuremath{\,\upmu\nospaceunit{m}}\xspace}

\def\fb   {\ensuremath{\aunit{fb}}\xspace}
\def\invfb   {\ensuremath{\fb^{-1}}\xspace}

\def\gsim{{~\raise.15em\hbox{$>$}\kern-.85em
          \lower.35em\hbox{$\sim$}~}\xspace}
\def\lsim{{~\raise.15em\hbox{$<$}\kern-.85em
          \lower.35em\hbox{$\sim$}~}\xspace}

\def\sPlot{\mbox{\em sPlot}\xspace}

\def\pt         {\ensuremath{p_{\mathrm{T}}}\xspace}

\def\ptot       {\ensuremath{p}\xspace}

\def\evtgen     {\mbox{\textsc{EvtGen}}\xspace}

\def\geant      {\mbox{\textsc{Geant4}}\xspace}

\def\photos     {\mbox{\textsc{Photos}}\xspace}

\def\pythia     {\mbox{\textsc{Pythia}}\xspace}

\def\tell1  {TELL1\xspace}
\def\ukl1   {UKL1\xspace}

\newcommand{\lhcborcid}[1]{\href{https://orcid.org/#1}{\hspace*{0.1em}\raisebox{-0.45ex}{\includegraphics[width=1em]{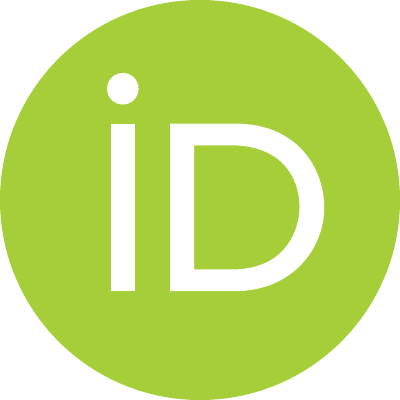}}}}

\newcommand{\rhoplus}{\ensuremath{\Prho^+}\xspace}
\newcommand{\piplus}{\ensuremath{\Ppi^+}\xspace}

\newcommand{\bctojpsirho}{\mbox{\ensuremath{\Bc\to\jpsi\rhoplus}}\xspace}
\newcommand{\bctojpsipi}{\mbox{\ensuremath{\Bc\to\jpsi\piplus}}\xspace}

\usepackage{cite} 
\usepackage{mciteplus}

\usepackage{longtable} 

\newenvironment{Tabular}[2][1]
  {\def\arraystretch{#1}\tabular{#2}}
  {\endtabular}

\begin{document}

\renewcommand{\thefootnote}{\fnsymbol{footnote}}
\setcounter{footnote}{1}


\begin{titlepage}
\pagenumbering{roman}

\vspace*{-1.5cm}
\centerline{\large EUROPEAN ORGANIZATION FOR NUCLEAR RESEARCH (CERN)}
\vspace*{1.5cm}
\noindent
\begin{tabular*}{\linewidth}{lc@{\extracolsep{\fill}}r@{\extracolsep{0pt}}}
\ifthenelse{\boolean{pdflatex}}
{\vspace*{-1.5cm}\mbox{\!\!\!\includegraphics[width=.14\textwidth]{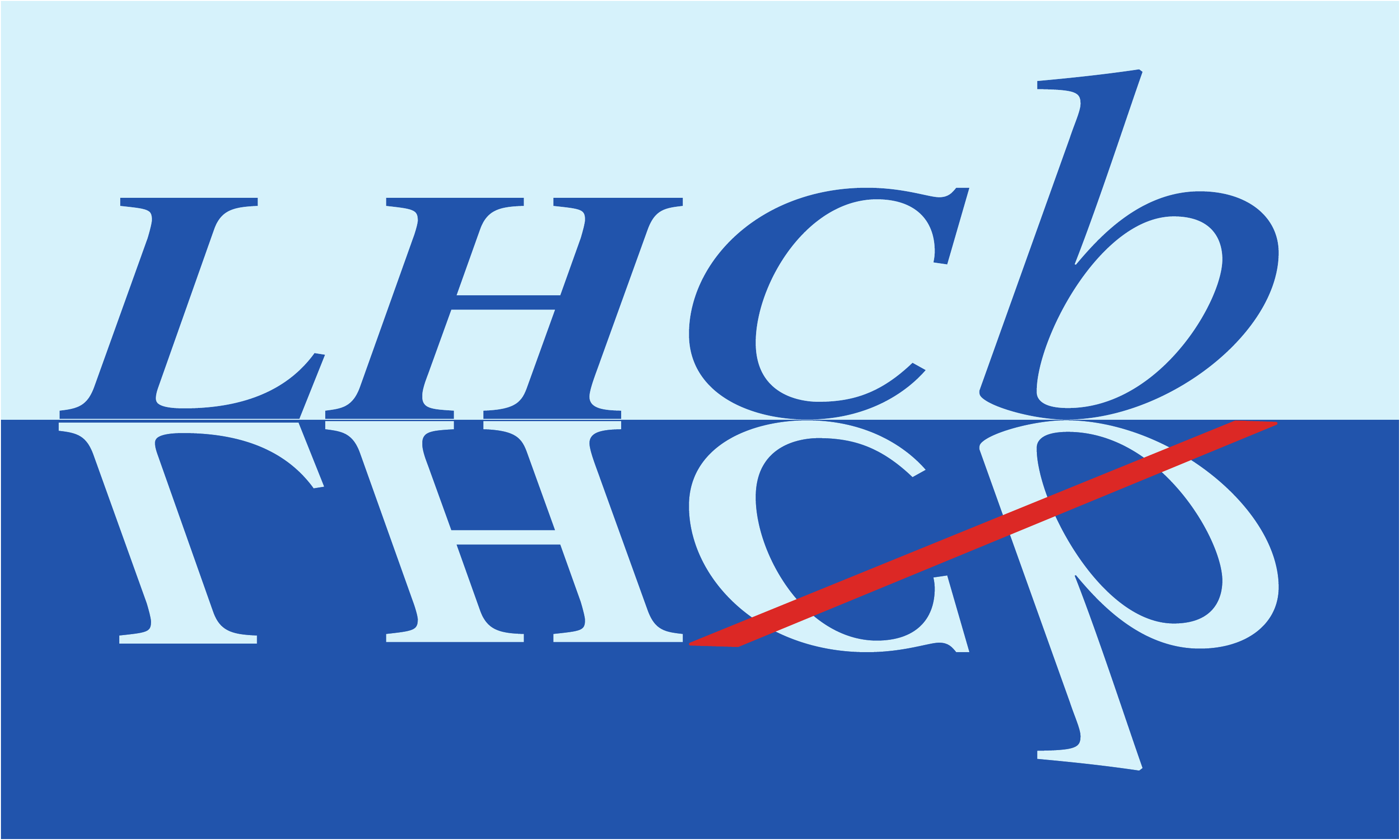}} & &}%
{\vspace*{-1.2cm}\mbox{\!\!\!\includegraphics[width=.12\textwidth]{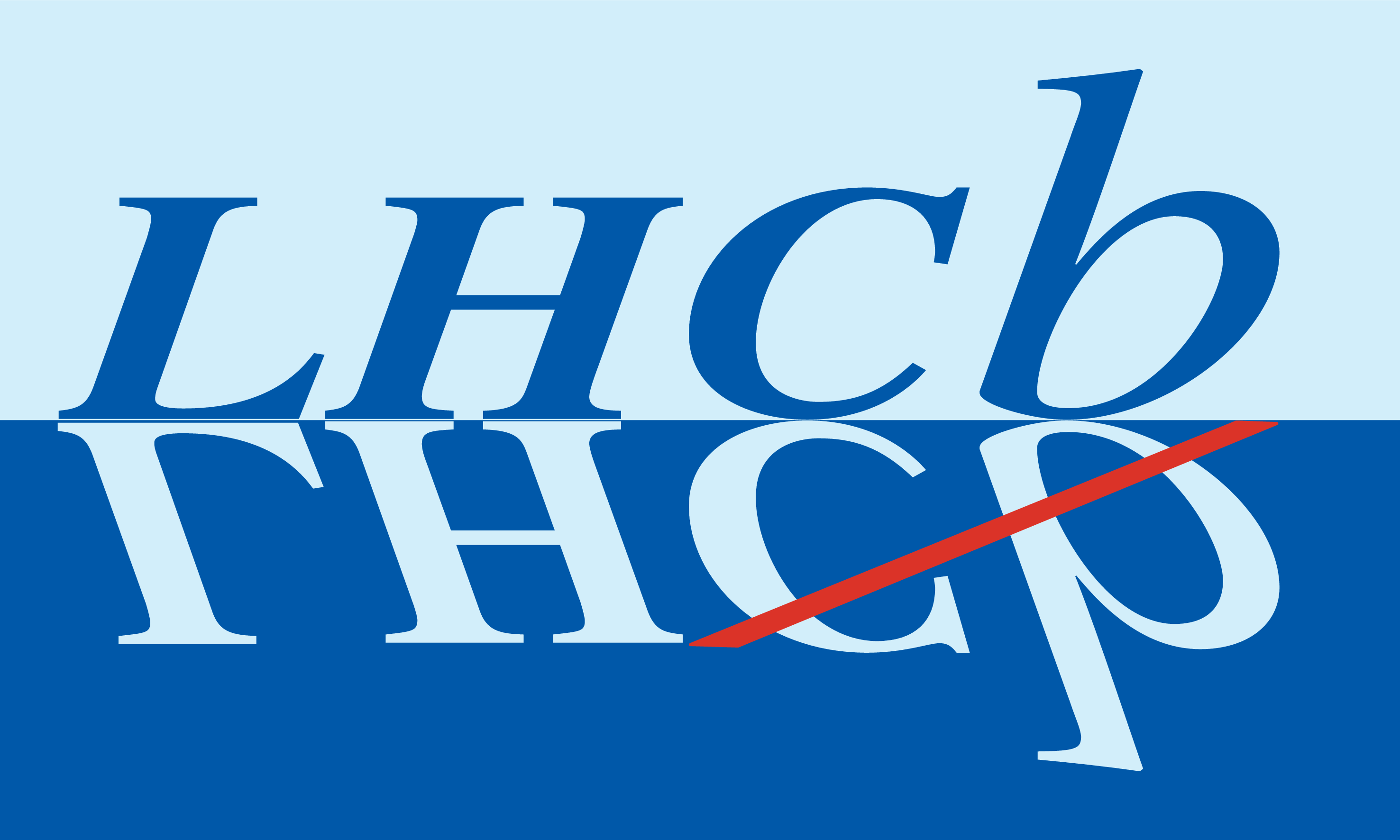}} & &}%
\\
 & & CERN-EP-2024-019 \\  
 & & LHCb-PAPER-2023-046 \\  
 & & February 6, 2024 \\ 
 & & \\
\end{tabular*}

\vspace*{2.0cm}

{\normalfont\bfseries\boldmath\huge
\begin{center}
  \papertitle 
\end{center}
}

\vspace*{2.0cm}

\begin{center}
\paperauthors\footnote{Authors are listed at the end of this paper.}
\end{center}

\vspace{\fill}

\begin{abstract}
  \noindent
  The~first observation of 
 the~$\decay{\Bc}{\jpsi\pip\piz}$~decay
  is reported
  with 
  high 
  significance 
  using proton\nobreakdash-proton collision data, 
  corresponding to an~integrated luminosity of 
  $9\invfb$,
  collected with the~\lhcb detector at 
  centre\nobreakdash-of\nobreakdash-mass energies of 7, 8, and 13\tev.
  The~ratio of its branching fraction relative to 
  the~$\decay{\Bc}{\jpsi\pip}$~channel is measured
  to~be 
  $$
    \dfrac{ \BR_{\decay{\Bc}{\jpsi\pip\piz}} }
          { \BR_{\decay{\Bc}{\jpsi\pip}} }
     =    2.80  \pm 0.15  \pm 0.11 \pm 0.16 \,,
  $$
  where the first uncertainty is statistical, 
 the second systematic and the third  
 related to imprecise knowledge of 
 the~branching fractions 
 for 
\mbox{$\decay{\Bu}{\jpsi\Kstarp}$}
 and \mbox{$\decay{\Bu}{\jpsi\Kp}$}~decays,
 which are used to determine the~\piz~detection efficiency.
 The~$\pip\piz$~mass spectrum is found 
 to be consistent with the~dominance of 
 an~intermediate \rhop~contribution
 in accordance with 
 a~model based on QCD~factorisation.
\end{abstract}

\vspace*{2.0cm}

\begin{center}
  Published in \href{https://doi.org/10.1007/JHEP04(2024)151}{JHEP04(2024) 151}
\end{center}

\vspace{\fill}

{\footnotesize 
\centerline{\copyright~\papercopyright. \href{\paperlicenceurl}{\paperlicence}.}}
\vspace*{2mm}

\end{titlepage}


\newpage
\setcounter{page}{2}
\mbox{~}
%
%
%
%


\renewcommand{\thefootnote}{\arabic{footnote}}
\setcounter{footnote}{0}

\cleardoublepage


\pagestyle{plain} 
\setcounter{page}{1}
\pagenumbering{arabic}


\section{Introduction}
\label{sec:introduction}
The~\Bc~meson, 
consisting of two heavy quarks of different flavours, 
was discovered by the~CDF collaboration~\cite{PhysRevLett.81.2432, 
PhysRevD.58.112004} at the~Tevatron collider, 
and is the~heaviest meson 
that can decay only through 
the~weak interaction. 
The~large $\bquark$\nobreakdash-quark production 
cross\nobreakdash-section 
at the~Large Hadron
Collider\,(LHC)~\cite{LHCb-PAPER-2010-002,
LHCb-PAPER-2011-003,
LHCb-PAPER-2011-043,
LHCb-PAPER-2013-004,
LHCb-PAPER-2013-016,
LHCb-PAPER-2015-037} 
enables the~ATLAS, CMS and \lhcb experiments 
to study in detail the~production, 
decays and other properties 
of the~\Bc~meson~\cite{LHCb-PAPER-2011-044, 
LHCb-PAPER-2012-054, 
LHCb-PAPER-2013-021, 
LHCb-PAPER-2013-047, 
LHCb-PAPER-2014-009, 
LHCb-PAPER-2021-034,
LHCb-PAPER-2022-025,
LHCb-PAPER-2015-024, 
LHCb-PAPER-2016-020, 
LHCb-PAPER-2013-010, 
LHCb-PAPER-2013-044, 
LHCb-PAPER-2013-063,
LHCb-PAPER-2014-039, 
LHCb-PAPER-2014-050, 
LHCb-PAPER-2014-060, 
CMS:2014oqy,
ATLAS:2015jep,
LHCb-PAPER-2016-001, 
LHCb-PAPER-2016-022,
LHCb-PAPER-2016-055, 
LHCb-PAPER-2016-058, 
LHCb-PAPER-2017-035, 
LHCb-PAPER-2017-045, 
CMS:2017ygm,         
CMS:2019uhm,         
LHCb-PAPER-2019-007, 
LHCb-PAPER-2019-033, 
LHCb-PAPER-2020-003, 
LHCb-PAPER-2021-023,
LHCb-PAPER-2023-037,
LHCb-PAPER-2023-039}.

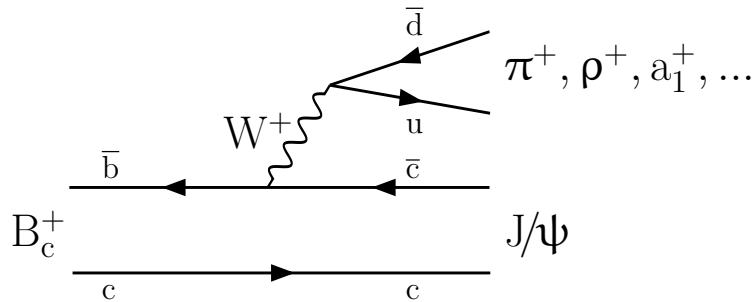
\begin{figure}[h]
  \setlength{\unitlength}{1mm}
  \centering
  \begin{picture}(90,45)
    \put(0,-5){
      \includegraphics*[width=70mm,height=50mm%
	  ]{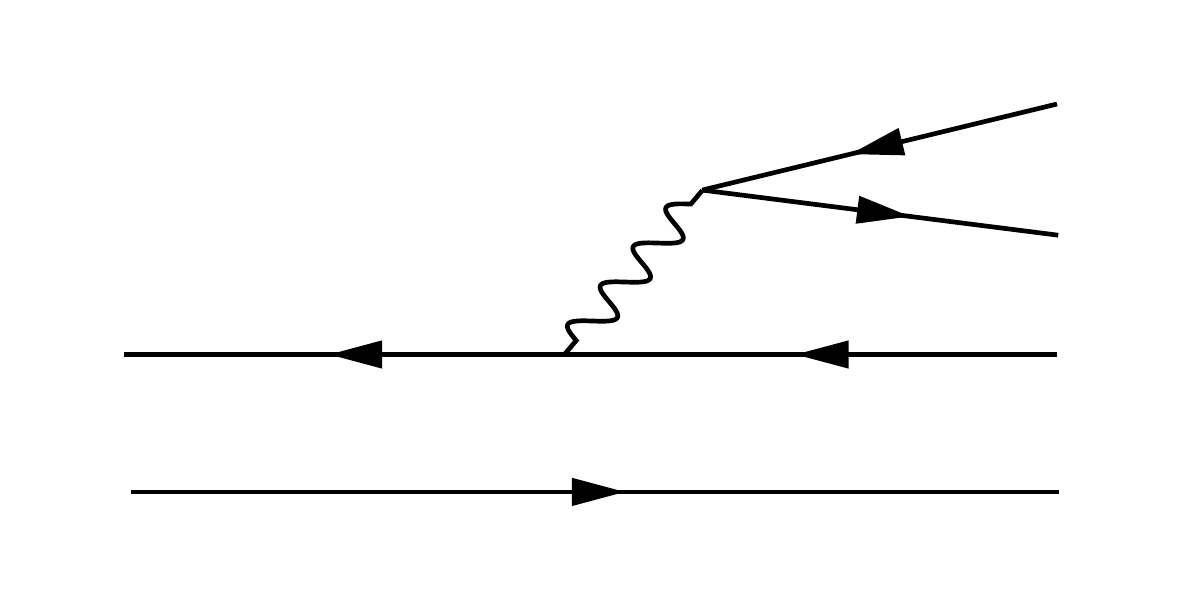}
    }
    \put(1 , 8){\Large\Bc}
    \put(66, 8){\Large\jpsi}
    \put(66,30){\Large$\pip,\rhoplus,\Pa_1^+,...$}
    \put(29,21){\Large$\PW^{+}$}
    \put(13,17){\bquarkbar}
    \put(13, 1){\cquark}
    \put(53,17){\cquarkbar}
    \put(53, 1){\cquark}
    \put(53,36){\dquarkbar}
    \put(53,23){\uquark}
  \end{picture}
  \caption {
   Diagram for the~decays of the~\Bc~meson into the~\jpsi~meson and light hadrons.
  }
  \label{fig:diagram}
\end{figure}

The~\Bc decay may proceed through the~weak decay of 
either of
its heavy constituents with the~other quark playing 
the~role of spectator,
or by a~$\bquarkbar\cquark$~annihilation into 
a~virtual $\PW^{+}$~boson.
The~lifetime of~the \Bc meson is about 
three times smaller~\cite{LHCb-PAPER-2013-063,
LHCb-PAPER-2014-060, 
CMS:2017ygm}
than that of the~\Bd~and \Bu~mesons~\cite{PDG2023}, 
confirming the~important
role played by the~charm quark in~\Bc~meson decays.
Its~decay to charmonium and 
light hadrons
proceeds via the~spectator diagram
as shown in Fig.~\ref{fig:diagram},
which is  
described
by the~quantum chromodynamics\,(QCD) 
factorisation approach~\cite{Bauer:1986bm,
Wirbel:1988ft}
using the~form factors of 
the~\mbox{\decay{\Bc}
{[\cquark\cquarkbar]\Wp}}~transition~\cite{Gershtein:1994jw,
Gershtein:1997qy,
Kiselev:1999sc,
Kiselev:2000pp,
Ebert:2002pp}
and the~universal spectral function 
for the~transition of the~virtual \Wp~boson 
into light hadrons~\cite{Likhoded:2009ib, 
Likhoded:2013iua, 
Berezhnoy:2011is,
Luchinsky:2012rk}.
This approach works remarkably well for
all~known decays of the~\Bc~meson
into charmonia with an~odd number of light 
hadrons in the~final state~\cite{LHCb-PAPER-2011-044,
      LHCb-PAPER-2012-054,
      LHCb-PAPER-2013-021,
      LHCb-PAPER-2013-047,
      LHCb-PAPER-2014-009,
      LHCb-PAPER-2015-024,
      LHCb-PAPER-2016-020,
      LHCb-PAPER-2021-034,
      LHCb-PAPER-2022-025}. 

The~\mbox{$\decay{\Bc}{\jpsi\pip\piz}$}~mode, 
which is the~simplest decay into a~charmonium 
state and 
an~even number of light hadrons, 
has not yet been observed. 
Its final state is expected to be dominated 
by the~\bctojpsirho mode with a~small admixture of 
\mbox{$\decay{\Bc}{\jpsi\Prho(1450)^+}$}~decays~\cite{Bruch:2004py}.
There are many predictions
for the~branching fraction ratio $\mathcal{R}$ of this decay 
relative to the~\mbox{$\decay{\Bc}{\jpsi\pip}$}~mode,
\begin{equation}
    \mathcal{R} \equiv 
    \dfrac{ \BR_{\decay{\Bc}{\jpsi\pip\piz}} }
          { \BR_{\decay{\Bc}{\jpsi\pip}} }\,,   \label{eq:R}
\end{equation}
which vary from 2.5 to 5.7~\cite{
Chang:1992pt, 
Liu:1997hr, 
Colangelo:1999zn,
AbdEl-Hady:1999jux,
Kiselev:2000pp, 
Kiselev:2001ej,
Ebert:2003cn,
Ivanov:2006ni,
Hernandez:2006gt,
Wang:2007sxa,
Likhoded:2009ib,
Qiao:2012hp,
Naimuddin:2012dy,
Kar:2013fna,
Rui:2014tpa,
Issadykov:2018myx,
Cheng:2021svx,
Zhang:2023ypl, 
Liu:2023kxr}. 
This
decay is an~important background
source for the~study of
other decays of 
\Bc~mesons~\cite{LHCb-PAPER-2023-039}
as well as 
rare decays of 
\Bd~mesons~\cite{LHCb-PAPER-2023-041}.

This paper reports on the~first observation of 
\mbox{$\decay{\Bc}{\jpsi\pip\piz}$}~decays\footnote{In 
this paper 
inclusion of charge-conjugate
decays is implied.} and the~measurement of 
$\mathcal{R}$.
A~sample of the~\mbox{$\decay{\Bc}{\jpsi\pip}$}~mode 
is used for normalisation
and a~high\nobreakdash-yield 
sample of~\mbox{$\decay{\Bu}{\jpsi\left(\decay{\Kstarp}
{\Kp\piz}\right)}$}~decays 
is used 
to~correct
the~detector resolution and any potential mass bias. 
The~analysis is 
performed using 
proton\nobreakdash-proton\,($\proton\proton$) collision data, 
corresponding to an~integrated 
luminosity of~9\,\invfb,
collected 
with the~\lhcb detector 
at~centre-of-mass energies of 7, 8, and 13\,\tev. 


\section{Detector and simulation}
\label{sec:detector}

The \lhcb detector~\cite{LHCb-DP-2008-001,LHCb-DP-2014-002} is 
a~single\nobreakdash-arm forward
spectrometer covering the~pseudorapidity range $2<\eta <5$,
designed for the study of particles containing \bquark or \cquark
quarks. 
The~detector includes a~high-precision tracking system
consisting of a~silicon\nobreakdash-strip vertex detector surrounding 
the~$\proton\proton$~interaction region~\cite{LHCb-DP-2014-001}, 
a~large-area silicon-strip detector located
upstream of a~dipole magnet with a bending power of about
$4{\mathrm{\,T\,m}}$, and three stations of silicon-strip detectors and straw
drift tubes~\cite{LHCb-DP-2013-003,LHCb-DP-2017-001}
placed downstream of the~magnet.
The~tracking system provides a~measurement of the~momentum \ptot 
of charged particles with
a~relative uncertainty that varies from 0.5\% at low momentum to 1.0\% at 200\gevc.
The~minimum distance of a track to a~primary $\proton\proton$~collision vertex\,(PV), 
the~impact parameter\,(IP), 
is measured with a~resolution of $(15+29/\pt)\mum$,
where \pt is the~component of the~momentum transverse to the beam in\,\gevc.
Different types of charged hadrons are distinguished using information
from two ring-imaging Cherenkov\,(RICH) detectors~\cite{LHCb-DP-2012-003}. 
Photons, electrons and hadrons are identified by a~calorimeter system consisting of
scintillating-pad and preshower detectors, an~electromagnetic
and a~hadronic calorimeter. 
Muons are identified by 
a~system composed of alternating layers of iron and multiwire
proportional chambers~\cite{LHCb-DP-2012-002}.

The online event selection is performed by a trigger~\cite{LHCb-DP-2012-004}, 
which consists of a~hardware stage, based on information from the~calorimeter and muon
systems, followed by a~software stage, 
which applies a~full event
reconstruction.
The~hardware trigger selects muon candidates 
with high transverse momentum 
or dimuon candidates with a~high value of 
the~product
of the~transverse momenta of the~two muons.
In~the~software trigger, 
two~oppositely charged muons are required to~form 
a~good\nobreakdash-quality
vertex that is significantly 
displaced from any~PV,
and the~mass of the~$\mumu$~pair 
is required to  
exceed~$2.7\gevcc$.

Simulated samples of 
\mbox{$\decay{\Bc}{\jpsi\pip\piz}$}, 
\mbox{$\decay{\Bu}{\jpsi\left(\decay{\Kstarp}{\Kp\piz}\right)}$}
and \mbox{$\decay{\Bc}{\jpsi\pip}$} decays
are used 
to model the~signal mass shapes, 
optimise the~selection requirements 
and compute the~efficiencies needed to determine 
the~branching fraction ratios.
The~\pythia~\cite{Sjostrand:2007gs} 
generator with a~specific
\lhcb configuration~\cite{LHCb-PROC-2010-056}
is used to simulate~\Bu~meson production 
from \proton\proton~collisions.
The~{\sc{BcVegPy}}~generator~\cite{Chang:2003cq,
Chang:2005hq,
Wang:2012ah,
Wu:2013pya}, 
which is 
interfaced to the~\pythia~parton shower and hadronisation model,
is used to simulate \Bc~meson production.
The~generator implements 
the~complete 
lowest\nobreakdash-order\,$(\upalpha_{\mathrm{s}}^4)$
perturbative QCD calculations
of the~dominant 
gluon\nobreakdash-gluon fusion 
\mbox{$\decay{\mathrm{gg}}{\Bc+ \cquarkbar + \bquark}$}
and 
\mbox{$\decay{\mathrm{gg}}
{\B^{*+}_{\cquark} + \cquarkbar + \bquark}$}~processes,
 neglecting the~contribution from 
 the~quark\nobreakdash-pair annihilation 
\mbox{$\decay{\mathrm{\quark\quarkbar}}
{\Bc+\cquarkbar + \bquark}$} and 
\mbox{$\decay{\mathrm{\quark\quarkbar}}
{ \B^{*+}_{\cquark}+ \cquarkbar + \bquark}$}~channels~\cite{Chang:1992jb,
Chang:1994aw,
Kolodziej:1995nv,
Chang:1996jt,
Berezhnoy:1996ks}.

Decays of unstable particles are described by 
the~\evtgen 
package~\cite{Lange:2001uf}, 
in which final\nobreakdash-state 
radiation is generated using \photos~\cite{davidson2015photos}. 
The ``BLL'' model 
for~\mbox{$\decay{\Bc}
{\jpsi\pip\piz}$}~decays~\cite{Likhoded:2009ib, 
Berezhnoy:2011nx,
Luchinsky:2012rk,
Likhoded:2013iua,
Berezhnoy:2011is}
assumes the~factorisation into $\Bc\to\jpsi\PW^{*+}$
with a~subsequent $\PW^{*+}\to\rhop$ transition
with a~tiny fraction 
of $\mathrm{W^{*+}}\to\Prho(1450)^+$
in accordance with
the~model by  K\"uhn and 
Santamaria~\cite{Kuhn:1990ad} and 
the~measurement of the~hadronic structure 
of the~\mbox{$\decay{\Ptau^+}{\pip\piz\neutb}$}~decay
by the~CLEO collaboration~\cite{CLEO:1999dln}.
The~interaction of the~generated particles 
with the~detector,
and its response, are implemented using
the~\geant 
toolkit~\cite{Allison:2006ve,*Agostinelli:2002hh} 
as described in Ref.~\cite{LHCb-PROC-2011-006}.
The~transverse momentum and rapidity spectra of 
the~\Bc mesons in simulated samples
are adjusted to match those observed 
in a~high\nobreakdash-yield, 
low-background sample of reconstructed
\mbox{$\decay{\Bc}{\jpsi\pip}$}~decays. 
The~detector response used for the~identification 
of pions and kaons 
is sampled from 
\mbox{$\decay{\Dstarp}{\left(\decay{\Dz}{\Km\pip}\right)\pip}$}
and \mbox{$\decay{\KS}{\pip\pim}$} calibration
channels~\cite{LHCb-DP-2012-003,
LHCb-DP-2018-001}. 
To~account for imperfections in the~simulation of
charged\nobreakdash-particle reconstruction, 
the~track\nobreakdash-reconstruction efficiency
determined from simulation 
is corrected using 
calibration samples
of \mbox{$\decay{\jpsi}{\mumu}$}~decays~\cite{LHCb-DP-2013-002}.
Samples of~\mbox{$\decay{\Bp}{\jpsi\left(\decay{\Kstarp}
{\Kp\left(\decay{\piz}{\g\g}\right)}\right)}$} decays 
are used to
correct the photon reconstruction efficiency 
in the~simulation~\cite{LHCb-PAPER-2012-022, 
LHCb-PAPER-2012-053, 
Govorkova:2015vqa,
Govorkova:2124605,
Belyaev:2016cri}.

\section{Event selection}
\label{sec:selection}

The~signal 
\mbox{$\decay{\Bc}{\jpsi\pip\piz}$}
and control
\mbox{$\decay{\Bu}{\jpsi\left(\decay{\Kstarp}
{\Kp\piz}\right)}$}~candidates 
are reconstructed using~\mbox{$\decay{\jpsi}{\mumu}$}~decays. 
The~same dimuon final state of the~\jpsi~meson
is used to reconstruct 
the~normalisation \mbox{$\decay{\Bc}{\jpsi\pip}$}~candidates.
A~loose preselection similar to that
used in Refs.~\cite{
LHCb-PAPER-2021-034,
LHCb-PAPER-2022-054,
LHCb-PAPER-2023-039} is applied,
followed by a~multivariate classifier to 
select a~higher purity subset of candidates.

Muons, pions and kaons are identified by
combining information from
the~RICH, calorimeter and muon detectors,
and they~are required to have transverse momenta
larger than 500\mevc.
Pions and kaons are further required to have a momentum 
between 3.2 and 150\gevc to ensure 
good performance 
of the~particle identification in 
the~RICH detectors~\cite{LHCb-PROC-2011-008,
LHCb-DP-2012-003}. 
To~reduce the combinatorial
background due to charged particles produced 
in \proton\proton~interactions, 
only tracks that are inconsistent with originating 
from any PV are used.
Photons are reconstructed from clusters 
in the~electromagnetic calorimeter
with transverse energy above~300\mev.
The~clusters must not be 
associated with reconstructed 
tracks~\cite{Terrier:691743,
LHCb-DP-2020-001}.

Pairs of oppositely charged 
muons consistent with originating from a~common vertex 
are combined to form 
\mbox{$\decay{\jpsi}{\mumu}$}~candidates. 
The~reconstructed mass of the~\mumu~pair 
is required to be
within the~range $3.0<m_{\mumu}<3.2\gevcc$. 
The~position of
the~reconstructed dimuon vertex is required to 
be separated from any reconstructed~PV.
The~\piz~candidates are reconstructed 
as
diphoton pairs with mass 
within $30\mevcc$~of the~known mass 
of the~\piz~meson~\cite{PDG2023}.

The~selected \mbox{$\jpsi\pip\piz$}, 
\mbox{$\jpsi\pip$} and
\mbox{$\jpsi\Kp\piz$}~combinations 
are used to form the~\Bc and \Bu~candidates.
The~$\pip\piz$ mass $m_{\pip\piz}$
is required to satisfy
\mbox{$620<m_{\pip\piz}<920\mevcc$}
and the~$\Kp\piz$~mass 
must be 
within~$50\mevcc$ of
the~known mass 
of the~\Kstarp~meson~\cite{PDG2023}. 
To~improve the~mass resolution
for the~\B~mesons, 
a~kinematic fit~\cite{Hulsbergen:2005pu} 
constrains 
the~masses of the~\jpsi and \piz~candidates 
to their known values~\cite{PDG2023}
and requires the~\B~candidates to originate from 
their associated PV.\footnote{Each 
\Bc or \Bu~candidate 
is associated with the~PV that 
yields the smallest $\chi^2_{\rm{IP}}$, 
which is defined 
as the~difference in the~vertex\nobreakdash-fit 
$\chi^2$ of a~given PV reconstructed with and
without the~particle under consideration.}
The~decay time 
$t_\B$
of 
the~\Bc\,(\Bu)~candidates
is required to satisfy 
\mbox{$0.1<t_\B<1.0\mm/c$} 
\mbox{$(0.2<t_\B<2.0\mm/c)$} 
to suppress 
the~large combinatorial 
background from tracks 
produced at the~PV
and from misreconstructed \B~candidates.
The~\mbox{$\decay{\Bc}{\jpsi\pip\piz}$}~candidates
with a~$\jpsi\pip$~mass within $45\mevcc$~of
the~known mass of the~\Bc~meson are vetoed 
to avoid contamination  
from \mbox{$\decay{\Bc}{\jpsi\pip}$}~decays 
with a~random \piz~added.
Similarly, candidates with
the~$\jpsi\pip$ mass 
between 5.180 and 5.305\gevcc 
are also vetoed to 
remove the~background
from~\mbox{$\decay{\Bu}{\jpsi\pip}$} 
and~\mbox{$\decay{\Bu}{\jpsi\Kp}$}~decays 
with a~random \piz~added.

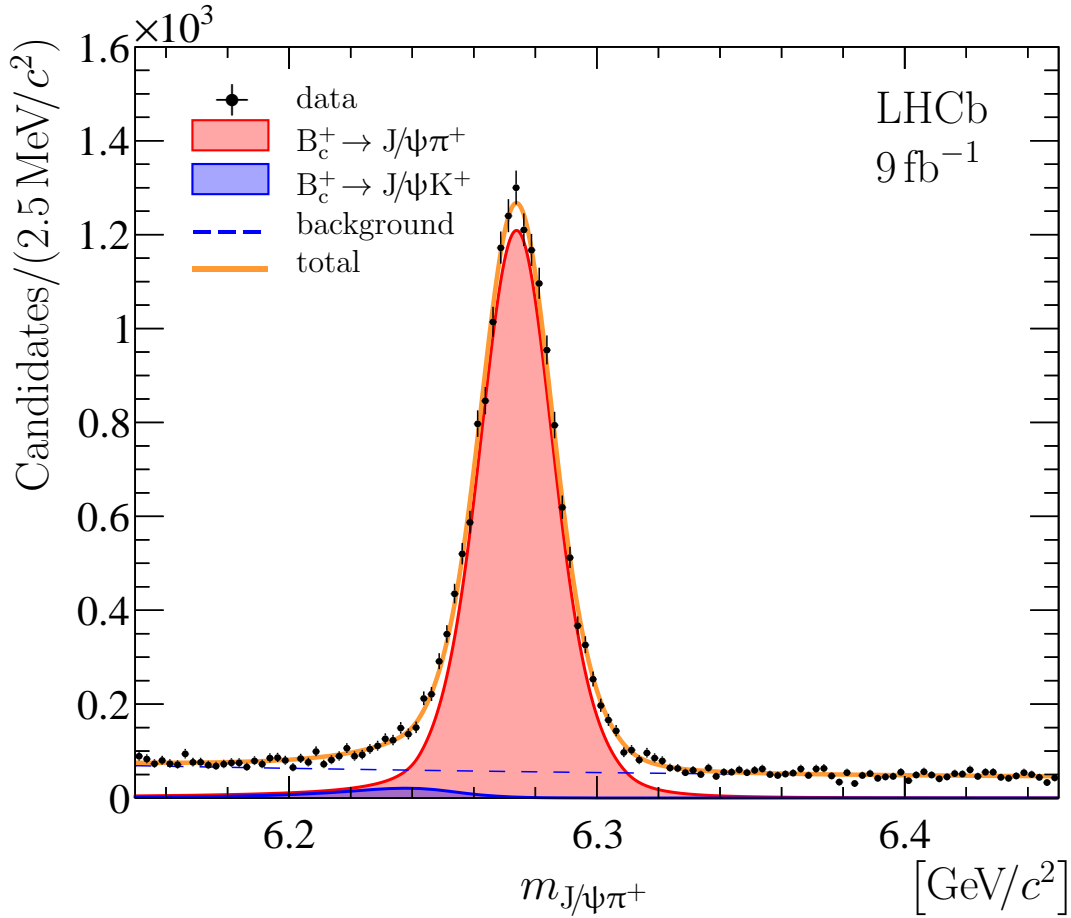
\begin{figure}[t]
  \setlength{\unitlength}{1mm}
  \centering
  \begin{picture}(150,120)
    \put(  0, 00){ 
      \includegraphics*[width=150mm,
	  ]{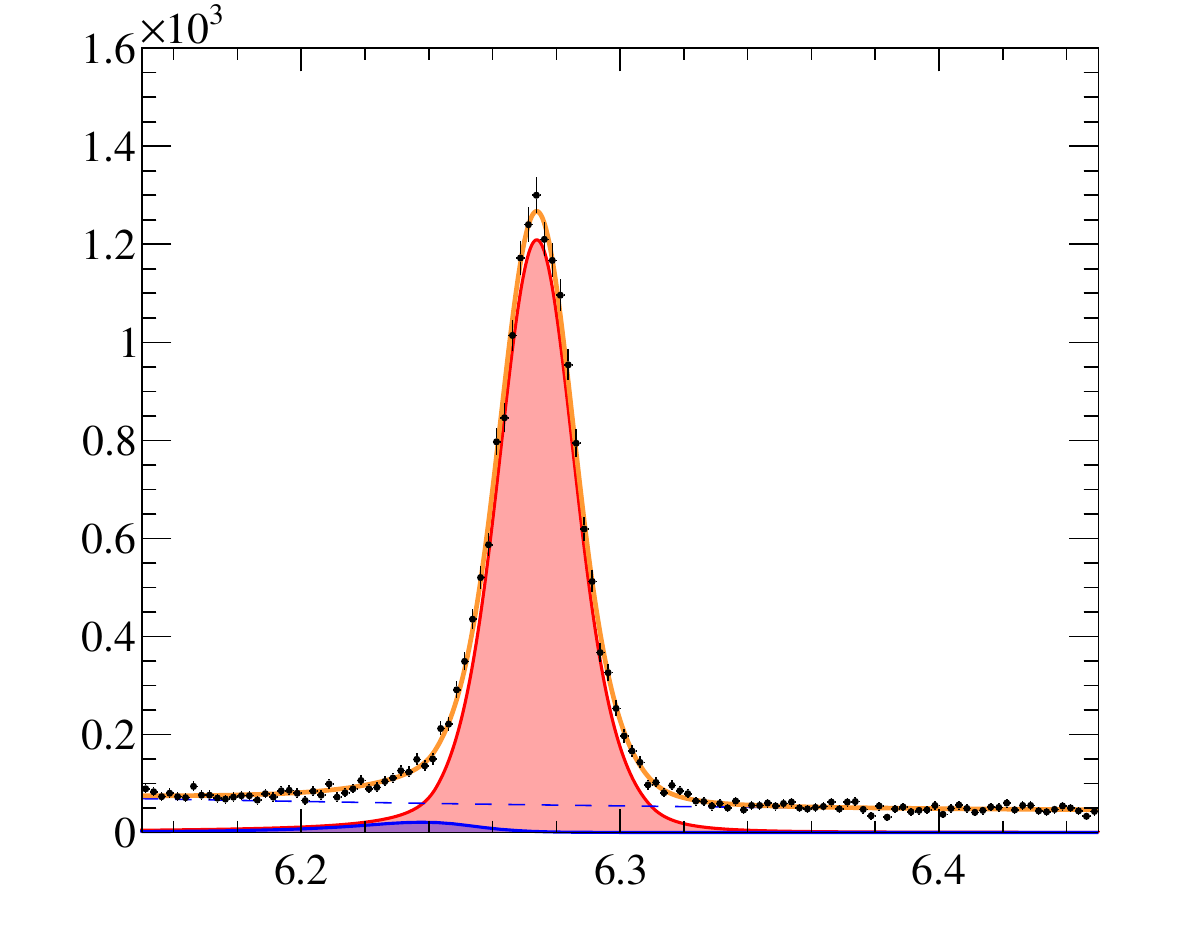}
      }  
      \put( 70, 1){\Large$m_{\jpsi\pip}$}
      \put(121, 1){\Large$\left[\!\gevcc\right]$}
      \put(3,55){\Large\begin{sideways}Candidates/(2.5\mevcc)\end{sideways}}
      \put(115,100){\Large$\begin{array}{l}\lhcb \\ 9\invfb\end{array}$}
	 \put(25,95){$\begin{array}{cl}
	 \!\!\!\bigplus\!\!\!\!\!\!\bullet\mkern-5mu&\mathrm{data}  
	 \\ 
	 \begin{tikzpicture}[x=1mm,y=1mm]\filldraw[fill=red!35!white,draw=red,thick]  (0,0) rectangle (10,4);\end{tikzpicture} & \decay{\Bc}{\jpsi\pip} 
	 \\
	 \begin{tikzpicture}[x=1mm,y=1mm]\filldraw[fill=blue!35!white,draw=blue,thick]  (0,0) rectangle (10,4);\end{tikzpicture} & \decay{\Bc}{\jpsi\Kp} 
	 \\
      {\color[RGB]{0,0,255}{\hdashrule[0.0ex][x]{10mm}{1.3pt}{2.3mm 0.8mm}}} & \mathrm{background}
	 \\
      {\color[RGB]{255,153,51} {\rule{10mm}{2.0pt}}} & \mathrm{total}
	 \end{array}$}
  \end{picture}
  \caption { \small
    Mass distribution
    for selected \mbox{$\decay{\Bc}{\jpsi\pip}$}~candidates
    with the~result of the~fit described in the~text.}
  \label{fig:rd_fits_pi_MLP}
\end{figure}

\begin{figure}[t]
  \setlength{\unitlength}{1mm}
  \centering
  \begin{picture}(150,120)
    \put(  0, 00){ 
      \includegraphics*[width=150mm,
	  ]{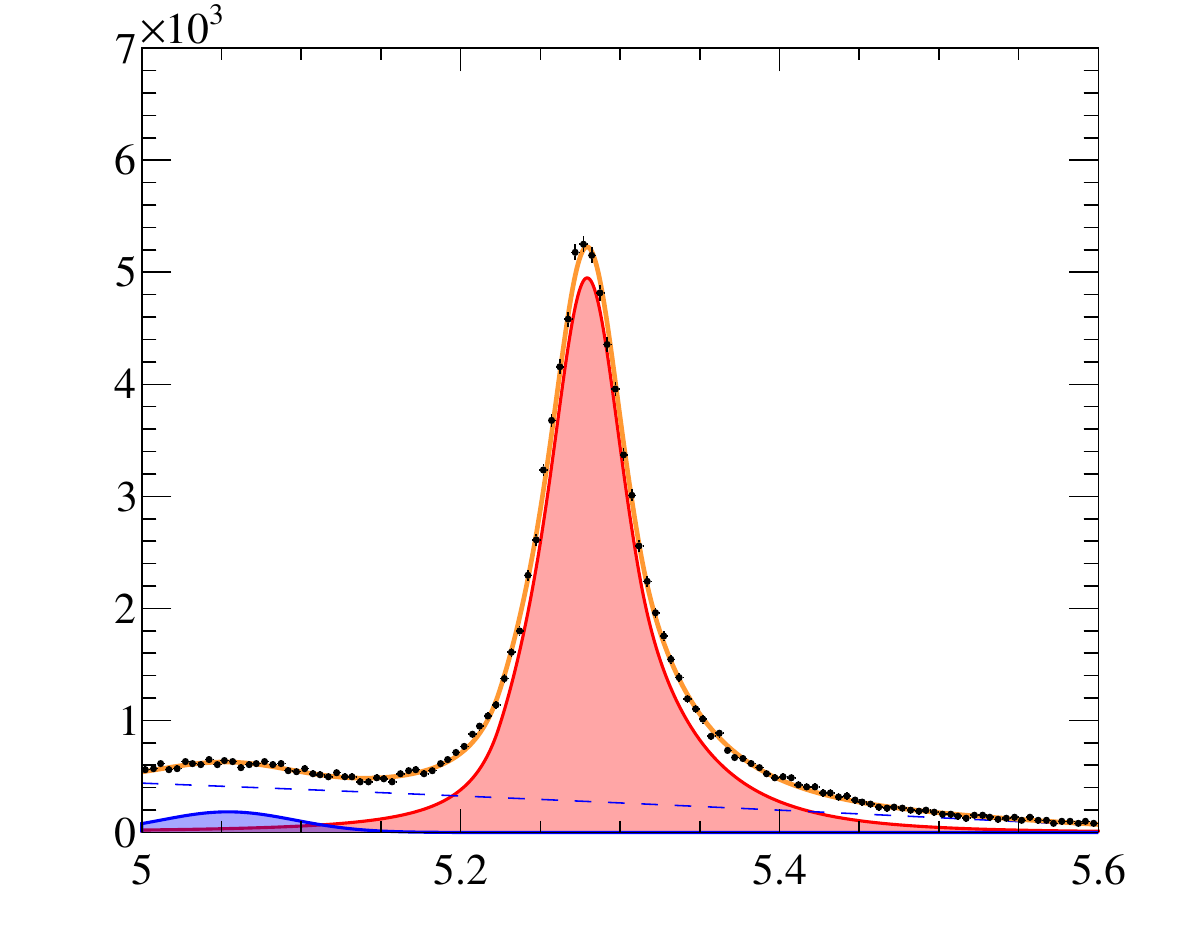}
      }  
    \put( 70, 1){\Large$m_{\jpsi\Kp\piz}$}
    \put(121, 1){\Large$\left[\!\gevcc\right]$}
    \put(5,58){\Large\begin{sideways}Candidates/(5\mevcc)\end{sideways}}
      \put(115,100){\Large$\begin{array}{l}\lhcb \\ 9\invfb\end{array}$}
	 \put(25,95){$\begin{array}{cl}
	 \!\!\!\bigplus\!\!\!\!\!\!\bullet\mkern-5mu&\mathrm{data}  
	 \\ 
	 \begin{tikzpicture}[x=1mm,y=1mm]\filldraw[fill=red!35!white,draw=red,thick]  (0,0) rectangle (10,4);\end{tikzpicture} & \decay{\Bu}{\jpsi\Kp\piz} 
	 \\
	 \begin{tikzpicture}[x=1mm,y=1mm]\filldraw[fill=blue!35!white,draw=blue,thick]  (0,0) rectangle (10,4);\end{tikzpicture} & \decay{\B}{\jpsi\Kp\piz\left(\Ppi\right)} 
	 \\
      {\color[RGB]{0,0,255}{\hdashrule[0.0ex][x]{10mm}{1.3pt}{2.3mm 0.8mm}}} & \mathrm{background}
	 \\
      {\color[RGB]{255,153,51} {\rule{10mm}{2.0pt}}} & \mathrm{total}
	 \end{array}$}
  \end{picture}
  \caption { \small
    Mass distribution
    for selected \mbox{$\decay{\Bu}{\jpsi\Kstarp}$}~candidates
    with 
    the~result of the~fit described 
    in the~text. }
  \label{fig:rd_fits_kst_MLP}
\end{figure}

\begin{figure}[t]
  \setlength{\unitlength}{1mm}
  \centering
  \begin{picture}(150,120)
    \put(  0, 0){ 
      \includegraphics*[width=150mm,
	  ]{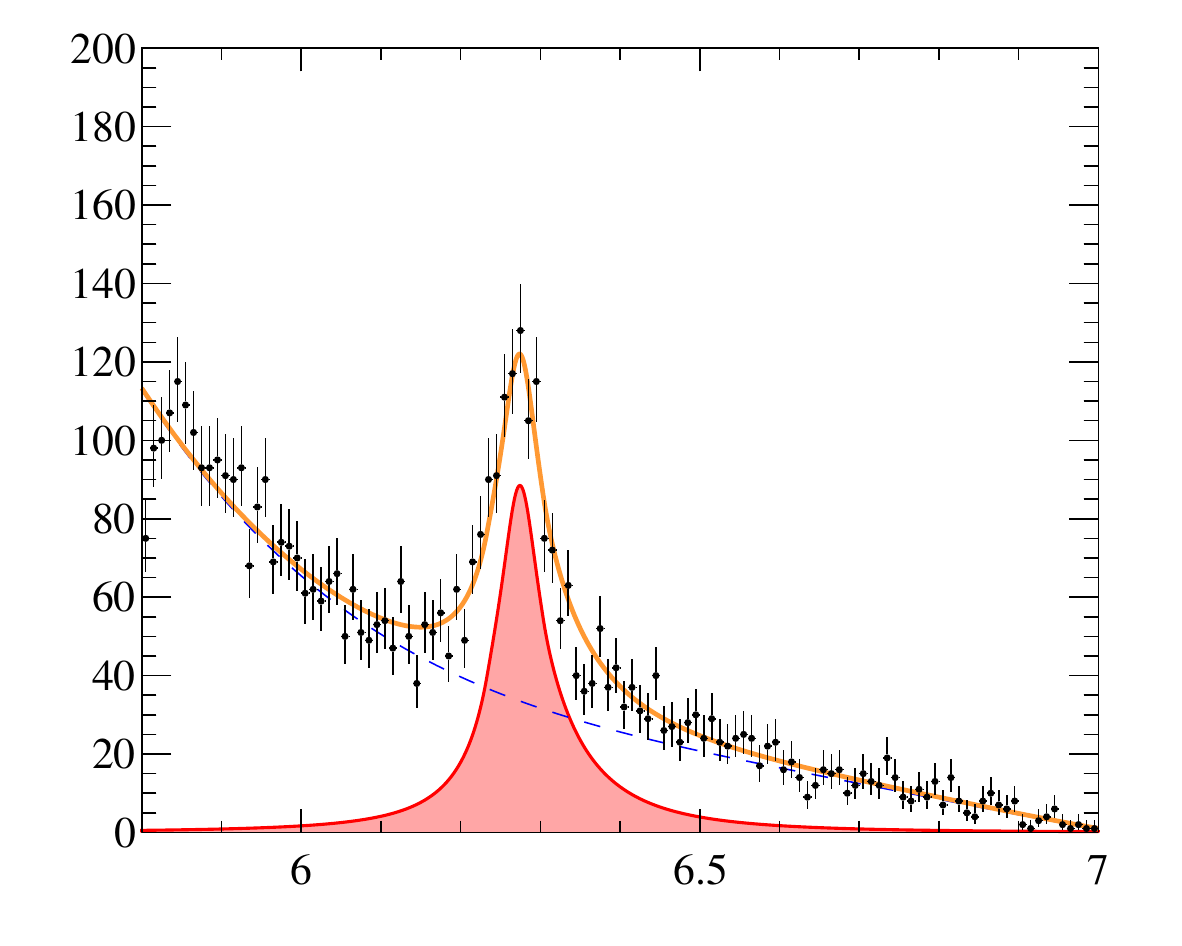}
      }  
      \put( 70, 1){\Large$m_{\jpsi\pip\piz}$}
      \put(121, 1){\large$\left[\!\gevcc\right]$}
    \put(1,56){\Large\begin{sideways}Candidates/(10\mevcc)\end{sideways}}
     \put(115,100){\Large$\begin{array}{l}\lhcb \\ 9\invfb\end{array}$}
	 \put(25,95){$\begin{array}{cl}
	 \!\!\!\bigplus\!\!\!\!\!\!\bullet\mkern-5mu&\mathrm{data}  
	 \\ 
	 \begin{tikzpicture}[x=1mm,y=1mm]\filldraw[fill=red!35!white,draw=red,thick]  (0,0) rectangle (10,4);\end{tikzpicture} & \decay{\Bc}{\jpsi\pip\piz} 
	 \\
    {\color[RGB]{0,0,255}{\hdashrule[0.0ex][x]{10mm}{1.3pt}{2.3mm 0.8mm}}} & \mathrm{background}
	 \\
      {\color[RGB]{255,153,51} {\rule{10mm}{2.0pt}}} & \mathrm{total}
	 \end{array}$} 
   \end{picture}
  \caption { \small
    Mass distribution
    for selected \mbox{$\decay{\Bc}{\jpsi\pip\piz}$}~candidates
    with 
    the~result of the~fit, described in the~text. 
    }
  \label{fig:rd_fits_rho_MLP}
\end{figure}

Further selection of
the~\Bc and \Bu~candidates
is based 
on a~multivariate estimator
known as a~multi\nobreakdash-layer 
perceptron\,({\sc{MLP}}) classifier.
The~{\sc{MLP}}~classifier 
is based on an~artificial neural 
network algorithm~\cite{McCulloch,rosenblatt58}
configured with a~cross\nobreakdash-entropy cost 
estimator~\cite{Zhong:2011xm}.
It~reduces the~combinatorial background to a~low level 
while retaining a~high signal efficiency. 
Three~{\sc{MLP}}~classifiers are trained separately for 
the~signal, normalisation and control candidates.
The~variables used 
in the~classifiers
include those 
related to the~reconstruction quality, 
kinematics and decay time of the~\Bc and \Bu~candidates, 
kinematics of the~final\nobreakdash-state 
hadrons and photons, 
as well as a~variable that characterises 
the~identification of charged 
pions and kaons. 
The~classifiers are trained using simulated 
signal samples, while  
the~\Bc and \Bu~candidates from data with mass
outside the~regions 
\mbox{$6.00< m_{\jpsi\pip\piz}< 6.50\gevcc$},
\mbox{$6.20< m_{\jpsi\pip}< 6.34\gevcc$} and 
\mbox{$5.10< m_{\jpsi\Kp\piz}< 5.50\gevcc$} 
are used to represent the~background for 
the~signal, normalisation and control modes, respectively. 
The~$k$\nobreakdash-fold
cross\nobreakdash-validation 
technique~\cite{chopping} with $k = 11$ 
is used to avoid 
introducing a~bias in the~{\sc{MLP}}~evaluation.  
The~requirement on each of the~{\sc{MLP}} classifiers 
is chosen to maximise 
the~figure\nobreakdash-of\nobreakdash-merit 
defined as 
$S/\sqrt{S+\kappa B }$,
where $S$~represents 
the~expected signal yield,
$B$~is the~background yield 
and $\kappa$~is
the~fraction of the~total background 
that affects the~signal determination.
The~background yield $B$ is calculated 
from fits to data, 
while $S=\varepsilon S_0$,
where $\varepsilon$ is the~efficiency of the~requirement 
on the~response of the~{\sc{MLP}}~classifier determined 
from simulation 
and $S_0$~is the~signal yield obtained 
from the~fit to 
the~data 
with a~loose 
requirement applied. 
The~factor $\kappa$~is determined from~pseudoexperiments. 
The~mass distributions 
for the~selected~\mbox{$\decay{\Bc}{\jpsi\pip}$},
\mbox{$\decay{\Bu}{\jpsi\Kstarp}$} and 
\mbox{$\decay{\Bc}{\jpsi\pip\piz}$}~candidates 
are shown in Figs.~\ref{fig:rd_fits_pi_MLP},
  \ref{fig:rd_fits_kst_MLP} and 
  \ref{fig:rd_fits_rho_MLP}, respectively.

\section{Determination of signal yields}
\label{sec:yields}

The~yields for 
the~normalisation
\mbox{$\decay{\Bc}{\jpsi\pip}$},
control 
\mbox{$\decay{\Bu}{\jpsi\Kstarp}$}
and signal 
\mbox{$\decay{\Bc}{\jpsi\pip\piz}$}~decays
are determined using 
extended unbinned maximum-likelihood fits to
the~$\jpsi\pip$,
$\jpsi\Kp\piz$ and 
$\jpsi\pip\piz$~mass spectra shown in 
Figs.~\ref{fig:rd_fits_pi_MLP},
  \ref{fig:rd_fits_kst_MLP} and 
  \ref{fig:rd_fits_rho_MLP}. 
 Each signal component is  parameterised 
 as the~sum of a~modified Gaussian function 
 with power law tails on both sides of 
 the~distribution~\cite{Skwarnicki:1986xj,
 LHCb-PAPER-2011-013} and a~standard 
 Gaussian function with common  
 location parameter.\footnote{The resulting signal shapes 
 are parameterised as a~location-scale family,  
 $f( \tfrac{x - \upmu}{\upsigma} ) $, using 
 the~location parameter $\upmu$ 
 and the~scale parameter $\upsigma$, 
 referred to as the~resolution parameter
 hereafter.}
 The~background components are parameterised 
 by positive decreasing second\nobreakdash-order 
 polynomials for the~fits 
 to the~$\jpsi\pip$ and $\jpsi\Kp\piz$~mass spectra
 and with a~positive convex decreasing 
 fourth\nobreakdash-order polynomial 
 for the~fit to 
 the~$\jpsi\pip\piz$~mass spectrum~\cite{karlin1953geometry}.
 In addition, the~fit model for the~$\jpsi\pip$~mass spectrum
 includes a~small component corresponding to 
 the~contribution from 
 Cabibbo\nobreakdash-suppressed 
 \mbox{$\decay{\Bc}{\jpsi\Kp}$}~decays~\cite{LHCb-PAPER-2013-021,
 LHCb-PAPER-2016-020} 
  where the charged kaon is reconstructed as a~pion. 
  The~shape of this component is taken from simulation. 
The~fit model for the~$\jpsi\Kp\piz$~mass spectrum 
includes an~additional~component corresponding to 
contributions from
\mbox{$\decay{\Bu}{\jpsi\Kp\piz\piz}$}
and~\mbox{$\decay{\Bd}{\jpsi\Kp\piz\pim}$} decays with 
missing $\piz$ or $\pim$~particles.
This component is parameterised using a~Gaussian function.
The~signal shape parameters, except 
for the~location parameters, 
are determined from simulation and their uncertainties 
are~propagated to the~fits using multivariate 
Gaussian constraints. 
For~each fit, the~location and 
resolution parameters are
expressed as 
\begingroup
\allowdisplaybreaks
\begin{subequations}
\begin{eqnarray}
    \upmu     & = & \updelta m_{\B} + m_\B \,,  
    \\ 
    \upsigma  & = & s \times \upsigma_{\mathrm{MC}}  \,,
\end{eqnarray}
\end{subequations}
\endgroup
where $m_\B$ is the~known mass of 
the~corresponding beauty meson~\cite{PDG2023}, 
$\upsigma_{\mathrm{MC}}$~is the~resolution parameter 
obtained from simulation, 
the~term~$\updelta m_{\B}$ corrects for a~possible 
mass bias, and the~scale 
factor $s$~takes into account
any difference in the~mass resolution 
between
simulation and data. 
The~parameters $\updelta  m_{\B}$ and $s$ 
are allowed to vary
in the~$\jpsi\pip$ 
and $\jpsi\Kp\piz$~mass spectra fits, 
while 
the~$\jpsi\pip\piz$~fit 
restricts 
them
to the~values obtained 
for the~\mbox{$\decay{\Bu}{\jpsi\Kp\piz}$}~mode 
with their corresponding uncertainties propagated 
using Gaussian constraints.

The~results of the~fits are shown in
Figs.~\ref{fig:rd_fits_pi_MLP},
\ref{fig:rd_fits_kst_MLP} and 
\ref{fig:rd_fits_rho_MLP}
with the~parameters 
summarised   
in Table~\ref{tab:fits}. 
The~statistical significance for 
the~\mbox{$\decay{\Bc}{\jpsi\pip\piz}$}~signal 
is estimated using Wilks' theorem~\cite{Wilks:1938dza}
and is found to exceed 20~standard deviations.  
To~validate the~observed 
\mbox{$\decay{\Bc}{\jpsi\pip\piz}$}~signal, 
several cross\nobreakdash-checks are performed. 
The~data are 
split into data\nobreakdash-taking periods with 
different polarity of 
the~dipole  magnet  
and into \Bcp and \Bcm~samples.
Alternative multivariate 
estimators, such as 
decision trees with gradient boosting~\cite{Breiman}
and a~projective likelihood estimator~\cite{Likelihood}, 
are used.
The~results are found to be consistent 
among all samples and analysis techniques.

  \begin{table}[b]
	\centering
	\caption{\small 
Yields $N$, mass biases 
 $\updelta m_\B$ and 
 resolution scale factors $s$ 
 from the~fits to 
 the~$\jpsi\pip$,
 $\jpsi\Kp\piz$ and $\jpsi\pip\piz$~mass spectra.
 The~uncertainties for $\updelta m_{\B}$ 
 include those from the~known
 masses of the~\Bc and \Bu~mesons.
 The~mass bias and resolution scale 
 parameters for the~\mbox{$\decay{\Bc}{\jpsi\pip\piz}$}~channel 
 are constrained by those from 
 the~\mbox{$\decay{\Bu}{\jpsi\Kp\piz}$}~control channel.
 }  
	\label{tab:fits}
	\vspace{2mm}
	\begin{tabular*}{0.80\textwidth}
	{@{\hspace{5mm}}l@{\extracolsep{\fill}}ccc@{\hspace{5mm}}}
 Channel 
 &  $N\ \left[10^3\right]$
 & $\updelta m_\B \ \left[\!\mevcc\right]$
 & $s$     
  \\[1.5mm]
  \hline 
  \\[-1.5mm]
\mbox{$\decay{\Bc}{\jpsi\pip}$}  
& $16.11\pm0.15$
& $-0.59\pm0.35$
& $1.10\pm0.01$ 
\\ 
\mbox{$\decay{\Bu}{\jpsi\Kp\piz}$}
& $81.91\pm0.52$
& $-0.03\pm0.20$
& $1.11\pm0.01$
\\
\mbox{$\decay{\Bc}{\jpsi\pip\piz}$} 
& $\phantom{0}1.08\pm0.06$ 
& $-0.11\pm0.49$
& $1.11\pm0.01$
   \end{tabular*}
\end{table}

\begin{figure}[t]
  \setlength{\unitlength}{1mm}
  \centering
  \begin{picture}(150,120)
    \put(  0, 0){ 
      \includegraphics*[width=150mm,
	  ]{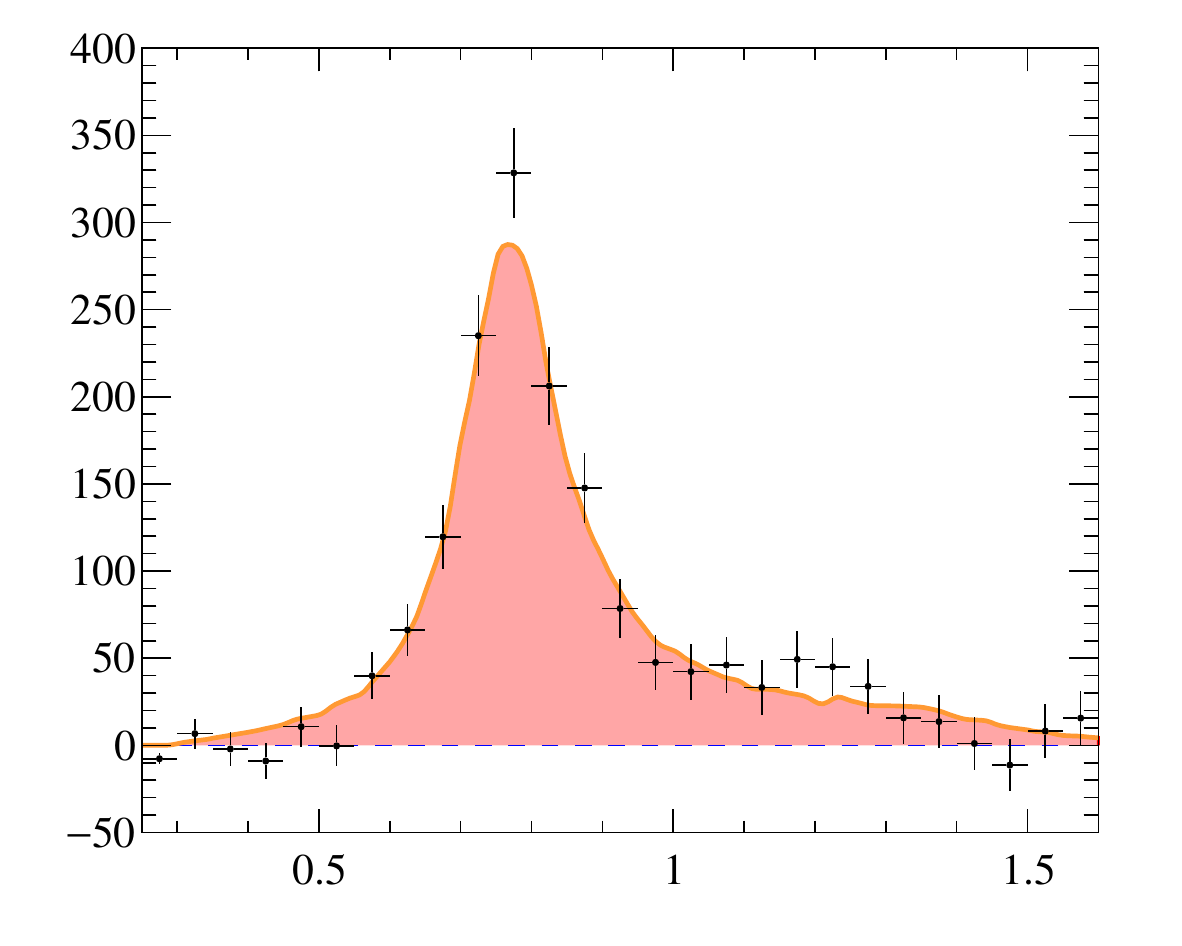}
     }  
     \put( 70, 1){\Large$m_{\pip\piz}$}
     \put(121, 1){\Large$\left[\!\gevcc\right]$}
     \put(1,70){\Large\begin{sideways}Yield/(50\mevcc)\end{sideways}}
     \put(25,100){\Large$\begin{array}{l}\lhcb \\ 9\invfb\end{array}$}
     \put(70,97){$\begin{array}{cl}
	 \!\!\!\bigplus\!\!\!\!\!\!\bullet\mkern-5mu&\mathrm{data}  
	 \\ 
	 \begin{tikzpicture}[x=1mm,y=1mm]\filldraw[fill=red!35!white,draw=red,thick]  (0,0) rectangle (10,4);\end{tikzpicture} 
      & \decay{\Bc}{\jpsi\pip\piz}\  {\mathrm{BLL}} 
	 \\
   {\color[RGB]{0,0,255}{\hdashrule[0.0ex][x]{10mm}{1.3pt}{2.3mm 0.8mm}}} 
     & \decay{\Bc}{\jpsi\left(\pip\piz\right)_{\text{P-wave,non-res.}}}
	 \\
    {\color[RGB]{255,153,51} {\rule{10mm}{2.0pt}}} & \mathrm{total}
	 \end{array}$} 
  \end{picture}
  \caption { \small
    Background-subtracted $\pip\piz$ mass distribution 
    from the~\mbox{$\decay{\Bc}{\jpsi\pip\piz}$}~decays.
    The~orange line shows the~result of 
    the~two\protect\nobreakdash-component
    fit described in the~text. 
    The~blue dashed line corresponds to 
    the~component parameterised by Eq.~\eqref{eq:nonres},
    which vanishes in the~fit.
    }
  \label{fig:pipi_MLP_BLL}
\end{figure}
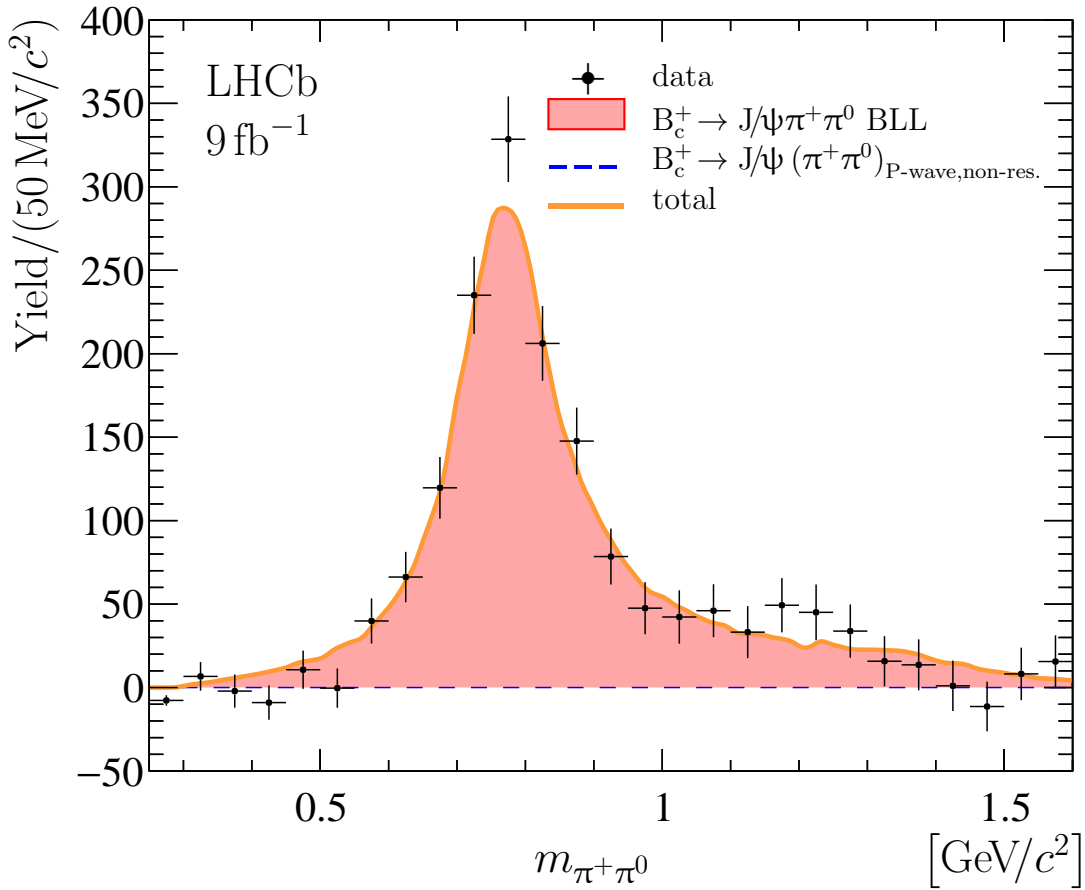

To validate the~predictions of the~BLL model,  
the~$\pip\piz$~mass spectrum from 
\mbox{$\decay{\Bc}{\jpsi\pip\piz}$}~decays 
is studied by extending the~$\pip\piz$~mass interval 
to the~region~$m_{\pip\piz}<1.6\gevcc$. 
The~background\nobreakdash-subtracted 
$\pip\piz$~mass spectrum 
is shown in Fig.~\ref{fig:pipi_MLP_BLL}, which
uses the~\sPlot~technique~\cite{Pivk:2004ty} 
based on the~fit to 
the~$\jpsi\pip\piz$~mass spectrum
with the~wider $\pip\piz$~mass interval. 
There is a~clear peak corresponding 
to the~decay $\rhop\to\pip\piz$, 
as expected by the~{\sc{BLL}}~model. 
To~quantify the~possible deviations from the~predictions, 
a~fit to the~$\pip\piz$~mass spectrum is 
performed using a~function consisting of 
two components.
 The~signal component 
 corresponds to the~coherent sum 
 of the~\mbox{$\decay{\Bc}{\jpsi\rhop}$}
 and \mbox{$\decay{\Bc}
{\jpsi}\Prho(1450)^+$}~contributions~\cite{Kuhn:1990ad},
  with the~shape obtained from 
  the~{\sc{BLL}}~model. 
 The~second component describes 
 contributions differing 
        from the~{\sc{BLL}}~model, generically parameterised as 
        \begin{equation}
            f(m_{\pip\piz}) \propto
            q^3(m_{\pip\piz})
             p(m_{\pip\piz}) \mathcal{P}^{+}_1(m_{\pip\piz})\,, \label{eq:nonres}
             \end{equation}
           where 
           $q$~is the~pion momentum in the~dipion 
           rest\nobreakdash-frame, 
           $p$~is the~$\pip\piz$~momentum in 
           the~\Bc~rest\nobreakdash-frame~\cite{Byckling}, 
           and $\mathcal{P}^+_1$ is a~positive first-order polynomial function~\cite{karlin1953geometry}.
           This~parameterisation accounts for the~phase\nobreak-space terms 
           and the~suppression of 
           the~S\nobreakdash-wave contribution. 
           The~unknown decay dynamics are absorbed 
           by the~polynomial function~$\mathcal{P}^{+}_1$. 
The result  of the~fit to~the
background\nobreakdash-subtracted 
$\pip\piz$~mass spectrum 
is superimposed 
in Fig.~\ref{fig:pipi_MLP_BLL}. 
The~contribution of the~second component vanishes in the~fit, 
demonstrating good agreement 
between the~observed $\pip\piz$~mass 
spectrum and the~expectations
from the~{\sc{BLL}}~model~\cite{Likhoded:2009ib, 
 Berezhnoy:2011nx,
 Luchinsky:2012rk,
 Likhoded:2013iua,
 Berezhnoy:2011is}.

\section{Branching fraction ratio computation}
\label{sec:ratio}

The~ratio of branching fractions
for the~\mbox{$\decay{\Bc}{\jpsi\pip\piz}$}
and 
\bctojpsipi~decays is
calculated as 
\begin{equation}
   \mathcal{R}= 
   \dfrac{N_{\decay{\Bc}{\jpsi\pip\piz}}}
         {N_{\decay{\Bc}{\jpsi\pip}}}
  \bigg/ 
  \dfrac{\varepsilon_{\decay{\Bc}{\jpsi\pip\piz}}}
       {\varepsilon_{\decay{\Bc}{\jpsi\pip}}}  \,,
\end{equation}
where the~yields 
${N_{\decay{\Bc}{\jpsi\pip\piz}}}$
and
${N_{\decay{\Bc}{\jpsi\pip}}}$
are taken from Table~\ref{tab:fits}, 
and $\varepsilon$~denotes 
the~corresponding efficiency, 
which is defined as the~product of 
the~detector acceptance
and the~reconstruction,
selection and trigger efficiencies, 
where each subsequent efficiency is defined with
respect to the~previous one. 
Each~of the partial efficiencies is calculated using 
the~simulation samples described in Sec.~\ref{sec:detector}.
The~ratio of branching fractions is found to be 
\begin{equation*} 
\mathcal{R}  =  
2.80 \pm 0.15 \,,
\end{equation*}
where the uncertainty is statistical only.

\section{Systematic uncertainties}
\label{sec:systematic}

Many sources of systematic uncertainties
cancel in the ratio of branching fractions $\mathcal{R}$
due to the~similarity between 
the~signal and normalisation decay channels.
The remaining contributions to the systematic uncertainty 
for $\mathcal{R}$ are summarised in 
Table~\ref{tab:systematics} 
and discussed below. 

\begin{table}[b]
	\centering
	\caption{\small
	Relative systematic uncertainties for
     the~ratio ${\mathcal{R}}$.
    The~total systematic uncertainty is 
    the~quadratic sum of individual contributions.
    The~systematic uncertainty of the~\piz~reconstruction
    efficiency due to 
    imprecise knowledge of 
    the~branching fraction of
    the~\mbox{$\decay{\Bu}{\jpsi\Kstarp}$}
    and  \mbox{$\decay{\Bu}{\jpsi\Kp}$}~decays
    is treated separately.
    } 
	\label{tab:systematics}
	\vspace{2mm}
	\begin{tabular*}{0.65\textwidth}
	{@{\hspace{5mm}}l@{\extracolsep{\fill}}c@{\hspace{5mm}}}
	Source  &   Uncertainty~$\left[\%\right]$ 
   \\[1.5mm]
  \hline 
  \\[-1.5mm]
  Fit model                           &     \\
  ~~~Signal shapes                    &   $\phantom{<}1.0$
  \\
  ~~~Background shapes                &   $\phantom{<}2.1$   
  \\
  Efficiency determination & 
  \\
  ~~~\Bc~kinematics                    &  $<0.1\,$      
  \\
 ~~~\Bc~decay model 
                                        &  $\phantom{<}2.2$
  \\
   ~~~Track reconstruction efficiency   &   $\phantom{<}0.1$
  \\
  ~~~\piz~reconstruction efficiency     &   $\phantom{<}0.9$ 
  \\
  ~~~Hadron identification             &   $\phantom{<}0.6$ 
  \\ 
  ~~~Trigger efficiency                &   $\phantom{<}1.1$ 
  \\
  Data-simulation difference           &   $\phantom{<}2.0$  
  \\
  Size of simulated samples            &   $\phantom{<}0.5$
    \\[1.5mm]
  \hline 
  \\[-1.5mm]
  Total                              &   $\phantom{<}4.1$
	\end{tabular*}
	\vspace{3mm}
\end{table}

An~important source of systematic uncertainty 
is the~imperfect knowledge 
of the~shapes of the~signal and background 
components used in the~fits. 
To~estimate this
effect, several alternative models 
are tested.  
The~\mbox{$\decay{\Bc}{\jpsi\pip\piz}$}~signal 
shape is parameterised as
a~sum of a~Gaussian
and three alternative
functions:
%
\begin{enumerate}
\item a~bifurcated generalised Student's $t$-distribution~\cite{Student,Jackman};
\item a~Johnson's $\mathcal{S}_{\mathrm{U}}$-distribution~\cite{JohnsonSU1,JohnsonSU2};
\item a~generalised hyperbolic distribution~\cite{GenHyp1,GenHyp2}.
\end{enumerate}
Similarly, the~\mbox{$\decay{\Bc}{\jpsi\pip}$}~signal
component is modelled as 
the~sum of a~Gaussian function
with three alternative shapes:
\begin{enumerate}
\item 
a bifurcated Student's $t$-distribution;
\item 
a~generalised hyperbolic distribution; 
\item 
a~Pearson's type~IV distribution~\cite{Pearson2}.
\end{enumerate} 
Five alternative background functions are tested for 
the~fit to the~$\jpsi\pip\piz$~mass spectrum:
\begin{enumerate}
\item 
a~positive decreasing convex third-order polynomial function~\cite{karlin1953geometry};
\item 
a~positive decreasing fourth-order polynomial function;
\item 
a~product of an~exponential function and a~positive third-order polynomial function;
\item 
a~sum of an~exponential function and 
  a~positive third-order polynomial function;
\item 
a~sum of a generalised Pareto distribution~\cite{Coles,Dargahi} 
and a~positive first\nobreakdash-order polynomial function.
\end{enumerate}
Lastly, for the~background component of the~$\jpsi\pip$~mass spectrum 
four alternative fit models are tested:
\begin{enumerate}
\item 
a~positive decreasing third-order polynomial function;
\item 
a~product of an~exponential function 
and a~positive first-order polynomial function; 
\item 
a~positive third-order polynomial function;
\item
a~positive fourth-order polynomial function.
\end{enumerate}
For~each alternative signal or background function, 
a~large ensemble of pseudoexperiments 
is produced and fit with the~baseline model.
The~absolute value of the~mean over an~ensemble 
for the~relative difference between 
the~fitted \mbox{$\decay{\Bc}{\jpsi\pip\piz}$}
or \mbox{$\decay{\Bc}{\jpsi\pip}$}~signal yields
and their baseline results is calculated.
Its~maximal value 
is found to be 1.0\%\,(2.1\%) for the~alternative 
signal\,(background) shapes 
and is taken as the~systematic uncertainty
related to the~fit model.
For~each model test, the~signal significance 
of the 
newly observed \mbox{$\decay{\Bc}{\jpsi\pip\piz}$} decay
is estimated using 
Wilks' theorem~\cite{Wilks:1938dza} and
is always found to exceed 20 standard deviations.

An uncertainty may originate
from possible differences in the~\Bc production kinematics between
data and simulation.
The~transverse momentum and rapidity spectra of 
the~\Bc mesons in the simulation are 
weighted
to match those observed in a~high\nobreakdash-yield, low\nobreakdash-background
sample of \mbox{$\decay{\Bc}{\jpsi\pip}$} decays. 
The~systematic uncertainty of the~efficiency ratio caused by 
the finite size of this sample 
is estimated by varying the \Bc-meson production kinematic spectra
within their uncertainties in the~weighting procedure.
The~induced variation for the~ratio 
$\mathcal{R}$
is found to be smaller than~0.1\%.

Simulated \mbox{$\decay{\Bc}{\jpsi\pip\piz}$}~events 
are corrected for polarisation of the~dipion system
in the~\Bc~rest frame 
to reproduce the~corresponding 
angular distribution observed in data.
Variations of this~correction within statistical 
uncertainties causes a~relative variation of 2.2\% for 
the~efficiency for~\mbox{$\decay{\Bc}{\jpsi\pip\piz}$}~decays, 
which is taken as the~systematic uncertainty
related to the~\Bc~decay model.

There are residual differences in 
the~reconstruction 
efficiency of charged\nobreakdash-particle tracks
that do not cancel completely in the~ratio 
due to slightly different kinematic distributions of
the~final\nobreakdash-state particles. 
The~track\nobreakdash-finding 
efficiencies obtained from simulated samples
are corrected using calibration channels~\cite{LHCb-DP-2013-002}. 
The~uncertainties related to the~efficiency
correction factors are propagated to the~ratios 
of the~total 
efficiencies using pseudoexperiments
and found to be 0.1\%
for  the~ratio $\mathcal{R}$.

The~difference of the~photon reconstruction efficiency 
between data and simulation is studied using 
large samples of 
\mbox{$\decay{\Bu}{\jpsi(\decay{\Kstarp}{\Kp(\decay{\piz}{\g\g}))}}$}
and \mbox{$\decay{\Bu}{\jpsi\Kp}$}~decays~\cite{LHCb-PAPER-2012-022, 
LHCb-PAPER-2012-053, 
Govorkova:2015vqa, 
Govorkova:2124605,
Belyaev:2016cri}.
The~uncertainty for 
the~photon efficiency corrections 
has three components: 
statistical, systematic 
and those related to the~imprecise knowledge 
of the~ratio of branching fractions for~\mbox{$\decay{\Bu}{\jpsi\Kstarp}$}
and~\mbox{$\decay{\Bu}{\jpsi\Kp}$}~decays.
The~statistical and systematic uncertainties 
are propagated  to the~ratio of 
the~total efficiencies using
pseudoexperiments and 
the~resulting 0.9\%
is taken as the~systematic uncertainty related to 
the~photon reconstruction.
The~uncertainty due to imprecise knowledge 
of external branching fractions
of 5.9\%~\cite{PDG2023} is treated 
separately.  

The~simulated detector response used for identification of pions
is resampled from the~control channels~\cite{LHCb-DP-2012-003,
LHCb-DP-2018-001}. 
The~systematic uncertainty obtained
through this procedure arises from the~kernel 
shape used in the~estimation of the~probability
density distributions. 
An~alternative combined 
response is estimated using a~modified 
kernel shape and 
the~efficiency
models are regenerated~\cite{LHCb-PAPER-2020-025, 
Poluektov:2014rxa}.
The~difference between the~two 
efficiency ratios of 0.6\% 
is taken as the systematic
uncertainty related to hadron identification.

The systematic uncertainty
related to the~trigger efficiency has been previously studied
by comparing the ratios of the trigger efficiencies in data and simulation using 
large samples of~\mbox{$\decay{\Bp}{\jpsi\Kp}$} and 
\mbox{$\decay{\Bp}{\psitwos\Kp}$}~decays~\cite{LHCb-PAPER-2012-010}.
Based on this comparison,
a~conservative estimate of 
1.1\% for the~relative difference 
between data and simulation is taken
as the~corresponding systematic uncertainty.

The~imperfect data description by simulation 
due to remaining effects not described above
is studied by varying the~{\sc{MLP}}~selection criteria
for the~normalisation decay \mbox{$\decay{\Bc}{\jpsi\pip}$}.
The~observed maximal difference between the~efficiency estimated using 
data and simulation does not exceed 2.0\%.
This value is taken as the~corresponding systematic uncertainty
for the~ratio 
$\mathcal{R}$.
The~last systematic uncertainty 
considered is 
due to the~finite size of the~simulated samples,
which amounts to~0.5\%.
The~total systematic uncertainty is 
calculated as the~quadratic sum of 
the~individual contributions.


\section{Results and discussion}
\label{sec:results}

\begin{figure}[t]
  \setlength{\unitlength}{1mm}
  \centering
  \begin{picture}(160,156)
    \put( 10,  4){ 
      \includegraphics*[height=150mm%
	]{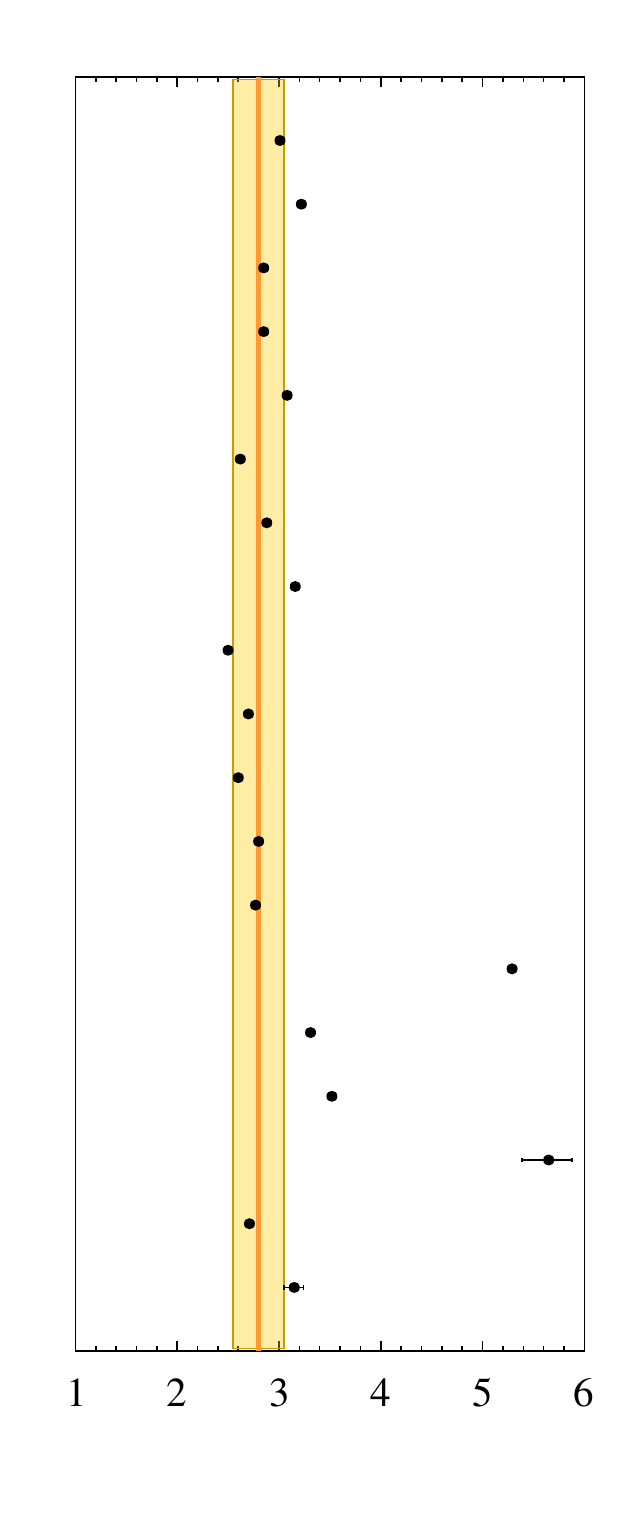}
    }
  \put(32,7){\large$\dfrac{\BR_{\decay{\Bc}{\jpsi\pip\piz}}}{\BR_{\decay{\Bc}{\jpsi\pip}}}$}
  \put(72,84){\scriptsize\begin{Tabular}[1.86]{lll} 
   Chang \& Chen                 & 1992        &  \cite{Chang:1992pt}       \\ 
   Liu \& Chao                   & 1997        &  \cite{Liu:1997hr}         \\
   Colangelo \& De Fazio         & 1999        &  \cite{Colangelo:1999zn}   \\
   Abd El-Hadi, Mu\~{n}oz \& Vary    & 1999        &  \cite{AbdEl-Hady:1999jux} \\
   Kiselev, Kovalsky \& Likhoded   & 2000        & \cite{Kiselev:2000pp, Kiselev:2001ej} \\
   Ebert, Faustov \& Galkin        & 2003        & \cite{Ebert:2003cn}       \\
   Ivanov, K\"orner \& Santorelli & 2006        &  \cite{Ivanov:2006ni}      \\
   Hern\'{a}ndez, Nieves \& Verde-Velasco &  2006 &  \cite{Hernandez:2006gt}   \\
   Wang, Shen \& Lu                & 2007        & \cite{Wang:2007sxa}     \\
   Likhoded \& Luchinsky         & 2009        &  \cite{Likhoded:2009ib}  \\
   Likhoded \& Luchinsky         & 2009        &  \cite{Likhoded:2009ib}     \\
   Likhoded \& Luchinsky         & 2009        &  \cite{Likhoded:2009ib}     \\ 
   Qiao {\it{et al.}}           &  2012       & \cite{Qiao:2012hp}        \\
   Naimuddin {\it{et al.}}      &  2012       & \cite{Naimuddin:2012dy,Kar:2013fna} \\
   Rui \& Zou                     & 2014        &  \cite{Rui:2014tpa}        \\
   Issadykov \& Ivanov            & 2018        &  \cite{Issadykov:2018myx}  \\
   Cheng {\it{et al.}}          & 2021        &  \cite{Cheng:2021svx}      \\
   Zhang                       & 2023        &  \cite{Zhang:2023ypl}       \\
   Liu                         & 2023        &  \cite{Liu:2023kxr}   
   \end{Tabular}}
  \end{picture}
  \caption { \small
  The measured ratio 
  of the~branching fractions 
  of the~\mbox{$\decay{\Bc}{\jpsi\pip\piz}$}
  and~\mbox{$\decay{\Bc}{\jpsi\pip}$}~decays 
in comparison with 
theory predictions.
The~colored band shows the~measured 
ratio with corresponding uncertainties
added in quadrature.}
  \label{fig:theory_exp}
\end{figure}

  The $\decay{\Bc}{\jpsi\pip\piz}$~decay
  is observed 
  with overwhelming significance 
  using proton\nobreakdash-proton collision data, 
  corresponding to an~integrated luminosity of 
  $9\invfb$,
  collected with the~\lhcb detector at 
  centre\nobreakdash-of\nobreakdash-mass energies of 7, 8, and 13\tev.
The~ratio of the~$\decay{\Bc}{\jpsi\pip\piz}$ branching fraction
relative to the~\mbox{$\decay{\Bc}{\jpsi\pip}$}~channel 
is measured to be: 
\begin{equation} \label{eq:result}
\mathcal{R} = 
   \dfrac{ \BR_{\decay{\Bc}{\jpsi\pip\piz}} }
         { \BR_{\decay{\Bc}{\jpsi\pip}} } 
  =  2.80 \pm 0.15 \pm 0.11 \pm 0.16 \,,
\end{equation}
where the first uncertainty is statistical, 
 the second systematic and the third 
 due to uncertainties 
in the~\mbox{$\decay{\Bu}{\jpsi\Kstarp}$}
 and \mbox{$\decay{\Bu}{\jpsi\Kp}$}
 branching fractions.
The~comparison of the~measured ratio
with theory predictions~\cite{Chang:1992pt, 
Liu:1997hr, 
Colangelo:1999zn,
AbdEl-Hady:1999jux,
Kiselev:2000pp, 
Kiselev:2001ej,
Ebert:2003cn,
Ivanov:2006ni,
Hernandez:2006gt,
Wang:2007sxa,
Likhoded:2009ib,
Qiao:2012hp,
Naimuddin:2012dy,
Kar:2013fna,
Rui:2014tpa,
Issadykov:2018myx,
Cheng:2021svx,
Zhang:2023ypl, 
Liu:2023kxr}
is shown in
Fig.~\ref{fig:theory_exp}.
While most of the~theory predictions
agree well with the~measurement,  
calculations from Refs.~\cite{Rui:2014tpa,
Issadykov:2018myx}
are only marginally consistent
and 
results from 
Refs.~\cite{Naimuddin:2012dy,Kar:2013fna,Cheng:2021svx}
are inconsistent with
the~measured value.
 The~study of the~$\pip\piz$~mass spectrum 
 indicates that the~\mbox{$\decay{\Bc}{\jpsi\pip\piz}$}~decay
 is saturated by 
 the~\mbox{$\decay{\Bc}{\jpsi\rhop}$}~contribution,
 with a~small admixture of 
 the~\mbox{$\decay{\Bc}{\jpsi\Prho(1450)^+}$}~decay,
 in accordance with theoretical expectations~\cite{Bruch:2004py,
 Likhoded:2009ib, 
 Berezhnoy:2011nx,
 Luchinsky:2012rk,
 Likhoded:2013iua,
 Berezhnoy:2011is}.

\section*{Acknowledgements}
%
%
\noindent 
This paper is dedicated to the~memory of our dear friend
and colleague Prof.~Yury~\mbox{Zaitsev}, who
made an important contribution to this analysis. 
We thank  A.~K.~Likhoded, X.~Liu and A.~V.~Luchinsky
for stimulating discussions on the~decays of the~\Bc~meson.
We express our gratitude to our colleagues in the CERN
accelerator departments for the excellent performance of the LHC. We
thank the technical and administrative staff at the LHCb
institutes.
We acknowledge support from CERN and from the national agencies:
CAPES, CNPq, FAPERJ and FINEP\,(Brazil); 
MOST and NSFC\,(China); 
CNRS/IN2P3\,(France); 
BMBF, DFG and MPG\,(Germany); 
INFN\,(Italy); 
NWO\,(Netherlands); 
MNiSW and NCN\,(Poland); 
MCID/IFA\,(Romania); 
MICINN\,(Spain); 
SNSF and SER\,(Switzerland); 
NASU\,(Ukraine); 
STFC\,(United Kingdom); 
DOE NP and NSF\,(USA).
We~acknowledge the computing resources that are provided by CERN, 
IN2P3\,(France), 
KIT and DESY\,(Germany), 
INFN\,(Italy), 
SURF\,(Netherlands),
PIC\,(Spain), 
GridPP\,(United Kingdom), 
CSCS\,(Switzerland), 
IFIN\nobreakdash-HH\,(Romania), 
CBPF\,(Brazil),
and Polish WLCG\,(Poland).
We~are indebted to the communities behind the multiple open-source
software packages on which we depend.
Individual groups or members have received support from
ARC and ARDC\,(Australia);
Key Research Program of Frontier Sciences of CAS, CAS PIFI, CAS CCEPP, 
Fundamental Research Funds for the~Central Universities, 
and Sci. \& Tech. Program of Guangzhou\,(China);
Minciencias\,(Colombia);
EPLANET, Marie Sk\l{}odowska\nobreakdash-Curie Actions, ERC and 
NextGenerationEU\,(European Union);
A*MIDEX, ANR, IPhU and Labex P2IO, and 
R\'{e}gion Auvergne\nobreakdash-Rh\^{o}ne\nobreakdash-Alpes\,(France);
AvH Foundation\,(Germany);
ICSC\,(Italy); 
GVA, \mbox{XuntaGal}, GENCAT, Inditex, InTalent and Prog.~Atracci\'on Talento, 
CM\,(Spain);
SRC\,(Sweden);
the~Leverhulme Trust, 
the~Royal Society
 and UKRI\,(United Kingdom).




\addcontentsline{toc}{section}{References}
\setboolean{inbibliography}{true}
\bibliographystyle{LHCb}
\bibliography{main,standard,LHCb-PAPER,LHCb-CONF,LHCb-DP,LHCb-TDR}

\newpage
\centerline
{\large\bf LHCb collaboration}
\begin
{flushleft}
\small
R.~Aaij$^{36}$\lhcborcid{0000-0003-0533-1952},
A.S.W.~Abdelmotteleb$^{55}$\lhcborcid{0000-0001-7905-0542},
C.~Abellan~Beteta$^{49}$,
F.~Abudin{\'e}n$^{55}$\lhcborcid{0000-0002-6737-3528},
T.~Ackernley$^{59}$\lhcborcid{0000-0002-5951-3498},
J. A. ~Adams$^{67}$\lhcborcid{0009-0003-9175-689X},
A. A. ~Adefisoye$^{67}$\lhcborcid{0000-0003-2448-1550},
B.~Adeva$^{45}$\lhcborcid{0000-0001-9756-3712},
M.~Adinolfi$^{53}$\lhcborcid{0000-0002-1326-1264},
P.~Adlarson$^{79}$\lhcborcid{0000-0001-6280-3851}
% ...

{\footnotesize \it

% ...
$^{36}$Nikhef National Institute for Subatomic Physics, Amsterdam, Netherlands\\
% ...
$^{45}$Instituto Galego de F{\'\i}sica de Altas Enerx{\'\i}as (IGFAE), Universidade de Santiago de Compostela, Santiago de Compostela, Spain\\
% ...
$^{49}$Physik-Institut, Universit{\"a}t Z{\"u}rich, Z{\"u}rich, Switzerland\\
% ...
$^{53}$H.H. Wills Physics Laboratory, University of Bristol, Bristol, United Kingdom\\
% ...
$^{55}$Department of Physics, University of Warwick, Coventry, United Kingdom\\
% ...
$^{58}$School of Physics and Astronomy, University of Glasgow, Glasgow, United Kingdom\\
$^{59}$Oliver Lodge Laboratory, University of Liverpool, Liverpool, United Kingdom\\
% ...
$^{67}$Syracuse University, Syracuse, NY, United States\\
% ...
$^{79}$Department of Physics and Astronomy, Uppsala University, Uppsala, Sweden, associated to $^{58}$\\
% ...
}
\end{flushleft}

\end{document}